\pdfoutput=1
\documentclass{jfm}
\usepackage{graphicx}
\usepackage{natbib}
\usepackage{amsmath}
\usepackage{color}
\ifCUPmtlplainloaded \else
  \checkfont{eurm10}
  \iffontfound
    \IfFileExists{upmath.sty}
      {\typeout{^^JFound AMS Euler Roman fonts on the system,
                   using the 'upmath' package.^^J}%
       \usepackage{upmath}}
      {\typeout{^^JFound AMS Euler Roman fonts on the system, but you
                   dont seem to have the}%
       \typeout{'upmath' package installed. JFM.cls can take advantage
                 of these fonts,^^Jif you use 'upmath' package.^^J}%
      }
  \else
  \fi
\fi

\ifCUPmtlplainloaded \else
  \checkfont{msam10}
  \iffontfound
    \IfFileExists{amssymb.sty}
      {\typeout{^^JFound AMS Symbol fonts on the system, using the
                'amssymb' package.^^J}%
       \usepackage{amssymb}%
       \let\le=\leqslant  \let\leq=\leqslant
       \let\ge=\geqslant  
      }{}
  \fi
\fi

\ifCUPmtlplainloaded \else
  \IfFileExists{amsbsy.sty}
    {\typeout{^^JFound the 'amsbsy' package on the system, using it.^^J}%
     \usepackage{amsbsy}}
    {\providecommand\boldsymbol[1]{\mbox{\boldmath $##1$}}}
\fi

\newcommand\Rey{\mbox{\textit{Re}}}  % Reynolds number
\newcommand\Stk{\mbox{\textit{St}}}  % Stokes number
\newsavebox{\astrutbox}
\sbox{\astrutbox}{\rule[-5pt]{0pt}{20pt}}

\title[Breakup of aggregates in turbulence]{Internal stresses and
  breakup of rigid isostatic aggregates in homogeneous and isotropic turbulence}

\author[J. De Bona, A.S. Lanotte and M. Vanni]%
{Jeremias De Bona$^1$, Alessandra S. Lanotte$^2$\break
and Marco Vanni$^1$
  \thanks{Email address for correspondence: marco.vanni@polito.it}}

\affiliation{ $^1$Dip. di Scienza Applicata e Tecnologia,
  Politecnico di Torino, Corso Duca degli Abruzzi 24, 10129 Torino,
  Italy
  \\[\affilskip] $^2$CNR-ISAC Istituto di Scienze dell'Atmosfera e del
  Clima, and INFN, Sez. Lecce, Str. Prov. Lecce-Monteroni, 73100
  Lecce, Italy }

\pubyear{????}
\volume{???}
\pagerange{XXX--YYY}
\date{?; revised ?; accepted ?. - To be entered by editorial office}
\begin{document}

\maketitle

\begin{abstract}
By characterising the hydrodynamic stresses generated by statistically 
homogeneous and isotropic turbulence in rigid aggregates, we estimate
theoretically the rate of turbulent breakup of colloidal aggregates
and the size distribution of the formed fragments. The adopted method
combines Direct Numerical Simulation of the turbulent field with a
Discrete Element Method based on Stokesian dynamics. In this way, not
only the mechanics of the aggregate is modelled in detail, but the
internal stresses are evaluated while the aggregate is moving in the
turbulent flow. We examine doublets and cluster-cluster isostatic
aggregates, where the failure of a single contact leads to the rupture
of the aggregate and breakup occurs when the 
tensile force at a contact exceeds the cohesive strength of the
bond. Due to the different role of the internal stresses, 
the functional relationship between breakup frequency and turbulence dissipation rate is very different in the two cases.
{  In the limit of very small and very large values, the frequency of breakup scales exponentially with the turbulence dissipation rate for doublets, while it follows a power law for cluster-cluster aggregates.}
For the case of large isostatic
aggregates it is confirmed that the proper scaling length for maximum
stress and breakup is the radius of gyration. The cumulative fragment 
distribution function is nearly independent of the mean turbulence 
dissipation rate and can be approximated by the sum of a small erosive component and a term that is quadratic with respect to fragment size.

\end{abstract}


\section{Introduction}\label{sec:introduction}
The dispersion of particles in a liquid, as performed in the
compounding step of materials processing, involves the breakup of
solid aggregates by hydrodynamic stresses. In addition, breakup plays an important
role in many aggregation processes, where solid particles
coagulate under the effect of mechanical stirring and the final size
distribution is often determined by the balance between the
aggregation of the smaller particles and the rupture of the larger
ones. Both processes are frequently conducted under intense agitation
and in this case the hydrodynamic stress required for breakup is
provided by turbulence.

In turbulent breakup the complex dynamics of turbulence interacts with the
intricate morphology of the aggregates, since it is
the redistribution of the hydrodynamic force over the structure of the
aggregate that may cause the stress to accumulate in some critical
locations and exceed the cohesive strength. Hence, it is not surprising
that, despite considerable efforts, a basic understanding of breakup
dynamics in turbulent flows is still lacking.

So far the methods for the study of breakup have proceeded along two
main lines. One approach aims at predicting accurately the
hydrodynamic stresses in the aggregate in simple flow fields, such as
shear or elongational flow. Initially these studies were performed on
highly idealised aggregates with spherical symmetry and uniform
porosity \citep{Adler_1979a}, or radially decreasing porosity
\citep{Sonntag_1987a}, and by using Brinkman's equation to model the
flow in the porous region. Subsequently 
\citet{Horwatt_1992b,Horwatt_1992a} made it apparent that the
simulation of the stress distribution had to be based on more
realistic reproductions of the aggregates, because weak points or local
heterogeneities in the structure, which are not captured by idealised
representations, play a fundamental role in starting breakup.
This consideration led to the application of Discrete
Element Methods (DEMs), which model in detail the disordered structure
of the aggregates and take into account the effect that each primary
particle has in generating and transmitting stress. Different
techniques were employed to estimate hydrodynamic forces in DEMs,
including free draining approximation
\citep{Potanin_1993,Becker_2009,Eggersdorfer_2010}, simplified
hydrodynamic screening models
\citep{Higashitani_1998,Higashitani_2001,Fanelli_2006a}, method of
reflections \citep{Gastaldi_2011} and Stokesian dynamics
\citep{Harada_2006,Vanni_2011,Seto_2011,Harsche_2012}. DEMs allow the
estimation of the strength of the flow field required to break a given
aggregate. Usually such results are summarised in terms of a breakup
exponent $m$, that relates the size of the aggregate, $d$, with the
critical shear rate needed to break it, $\dot\gamma_{cr}$, and that was found
to depend on aggregate morphology \citep{Zaccone_2009}:
\begin{equation}
d \propto \dot{\gamma}_{cr}^{-m}\,.
\label{eq:fracture_exponent}
\end{equation}
Empirical studies and semi-theoretical analyses based on simplified
energy balances
\citep{Parker_1972,Kobayashi_1999,Bache_2004,Wengeler_2007} suggest
that eq.~(\ref{eq:fracture_exponent}) also estimates the size of the 
aggregates resulting from a process of turbulent breakup, provided
that the effective mean shear rate for the turbulent flow is used:
\begin{equation}
  \dot{\gamma}_{\rm eff} = \sqrt{{\left< \varepsilon \right>}/\nu}\,,
  \label{eq:eff_shear_rate}
\end{equation}
where $\left< \varepsilon \right>$ is the mean kinetic energy
dissipation rate and $\nu$ is the fluid kinematic
viscosity. 

While laminar shear flows are
characterised by a single value of shear rate, in a turbulent flow
$\dot\gamma_{\rm eff}$ is the outcome of a wide distribution of
instantaneous and local shear rates. Hence, the occurrence of a value
of such shear rate strong enough to induce breakup obeys a statistical
spatial and temporal distribution dictated by the turbulence. In order 
to take this effect into account, the study of turbulent breakup normally followed a
different approach, which was aimed at the prediction of the breakup
rate and did not consider the details of the structure of the
aggregates.

The rate of turbulent breakup for particles of a given size $d$ is
the number of such particles that undergo rupture per unit time and
volume of suspension. It can be expressed as $f_{br}(d) n(d)$, where
$n(d)$ is the number concentration of the particles and $f_{br}(d)$ is
called breakup frequency or, in the field of population balances,
breakup kernel \citep{Marchisio_2013,Vanni_2000b}. The simplest models
for the breakup frequency \citep{Lu_1985,Pandya_1983,Spicer_1996c} 
prescribe a power law relationship in terms of shear rate and aggregate size:
\begin{equation}
  f_{br}(d) \propto \dot{\gamma}_{\rm eff}^\alpha d^\beta\,,
  \label{eq:powerlaw_br_frequency}
\end{equation}
where $\alpha$ and $\beta$ are positive empirical parameters that
account for the fact that breakup is faster for larger aggregates and
stronger turbulence. A more theoretical approach was developed by
\citet{KustersPhD_1991}, who assumed that in a homogeneous turbulent
system, where the hydrodynamic field is continuously changing in time,
breakup occurs whenever the local instantaneous velocity gradient
around the aggregate exceeds the critical value given by
eq.~(\ref{eq:fracture_exponent}). Hence, the breakup rate is
determined by the frequency at which the shear rate becomes larger
than $\dot\gamma_{cr}$ \citep{Delichatsios_1975}. By assuming a normal
distribution for the local shear rate \citep{Delichatsios_1976} the
breakup frequency of small particles results in the following
expression:
\begin{equation}
f_{br}(d) = \sqrt{\frac{4\,\left< \varepsilon \right>}{15\pi\nu}} \,
\exp{ \left[ -\frac{15}{2} \frac{\nu}{\left< \varepsilon \right>}
    [\dot\gamma_{cr}(d)]^2 \right] }\,.
\label{eq:kusters_br_frequency}
\end{equation}

A similar point of view was adopted by B\"abler, who characterised
turbulence by a multifractal model \citep{Baebler_2008a} and proposed
a breakup rate function that is equal to
eq.~(\ref{eq:kusters_br_frequency}) for small aggregates, while it
becomes a power law in the limit of very large ones. In a subsequent work \citet{Baebler_2012} adopted a
first passage time statistics \citep{Redner_2001} to estimate the breakup rate by means of
Direct Numerical Simulation (DNS) data. Moreover,
they showed that an Eulerian proxy, based on the joint statistics of
the instantaneous energy dissipation and its time derivative, can
alternatively be used since it is easier to measure in experiments.

The previous approaches consider separately the roles of turbulence
and aggregate mechanics: aggregate mechanics and the behaviour in
simple flow fields set the relationship between aggregate size and
critical strain rate, $\dot\gamma_{cr}(d)$, whereas turbulence
determines the functional relationship between the breakup frequency
of an aggregate and its critical strain rate,
$f_{br}(\dot\gamma_{cr})$. However, the ability of the local flow
field to break an aggregate may depend not only on the strength of the
flow $\dot\gamma$, but also on the instantaneous orientation of the
particle with respect to the flow field and hence on the history of
the aggregate. A certain value of $\dot\gamma$ can break an aggregate
if its orientation enhances internal stresses, but may not be
sufficient in the case of unfavourable alignment. This implies an
interaction between mechanic and fluid dynamic aspects, and considering
them separately, as done in the aforementioned approaches, may not be
accurate, particularly in the case of small aggregates.

{ 
The method
used here combines DNS of the turbulent field with a DEM based on
Stokesian dynamics, which models in detail the mechanics of the
aggregate and evaluates internal stresses while the aggregate is
moving in the turbulent field. The adopted DEM approach is a refinement of the method used previously by Vanni and coworkers \citep{Vanni_2011,Sanchez_2012} for steady laminar flow fields. 

The breakup frequency is estimated by using the approach devised by \citet{Baebler_2012} to process data from the DNS of turbulence. However, the method by B\"abler does not take into account the effect of the structure on the redistribution of the internal stresses inside the aggregates and assumes that an aggregate breaks up whenever its critical flow strength $\dot\gamma_{cr}$ is exceeded. On the contrary, in our case breakup is assumed to occur when the internal stress on the most loaded intermonomer bond exceeds the cohesive force of the bond. In this way the effect of aggregate orientation on stress distribution and, consequently, on breakup is implicitly taken into account.

So far, simulations that capture in detail both the
fluid dynamic and the mechanical aspects of the problem for a turbulent system have been
restricted to doublets, that is, rigid aggregates formed by two contacting spherical monomers. 
In \citet{Derksen_2008}, the problem 
was solved numerically for particles with size sligthly larger than
the Kolmogorov length scale by calculating very precisely the flow
field around the particles with a lattice-Boltzmann method. Such an
approach is not practical for studying aggregates made by many
monomers or smaller than the Kolmogorov length scale, due to the
exceedingly high computational cost of the method. From this point of
view, the combination of DNS and Stokesian dynamics, although less
accurate, provides an excellent alternative and allowed us to
investigate the behaviour of aggregates that are more typical of the
colloidal domain.
}

In the first part of the work, we discuss the results obtained in the
simulation of doublets, where the physical interpretation of the
simulation is easier because of the simple geometry. In the second
part, we present the results for more realistic rigid cluster-cluster
aggregates of low fractal dimension made by several hundreds of
primary particles. Throughout the work, we consider isostatic
aggregates, that is clusters without redundant bonds, where the
failure of a single contact always leads to the breakup of the
structure. At a physical level, aggregates with such features are
normally obtained by aggregation of highly destabilised colloidal
suspensions, in which most collisions lead to permanent bonds and no
restructuring of the structure has occurred \citep{Gastaldi_2011}.

The two aforementioned cases of doublets and large cluster-cluster
aggregates are the extreme conditions of a range of situations. As
shown in the following sections, the doublet is the system for
which the effect of orientation is highest, since the single contact of this configuration is
subject to alternating traction and compression. Differently, this
effect is significantly reduced with big
cluster-cluster aggregates, where there are bonds under traction under any possible orientation, due to their large and disordered structure.

\section{Methods}\label{sec:methods}

\subsection{Main assumptions}
We investigate the breakup of small aggregates induced by stationary
homogeneous and isotropic turbulence in dilute suspensions. In this
case the finest structures of turbulence can be characterised by
Kolmogorov length scale $\eta_k$ and time scale $\tau_k$, defined as:
\begin{equation}
  \eta_k = \left( \nu^3/ \left< \varepsilon \right> \right)^{1/4}\,,
  \label{eq:kolmogorov_length}
\end{equation}
\begin{equation}
  \tau_k = \left( \nu / \left< \varepsilon \right> \right)^{1/2}\,.
  \label{eq:kolmogorov_time}
\end{equation}
We assume that the aggregates are much smaller than the scale of the
smallest turbulent motions, $(d/\eta_k) \ll 1$. This condition implies
that the aggregate Reynolds number $Re_p$ is very small too: 
\begin{equation}
  \Rey_p = \frac{\dot{\gamma}_{\rm eff} d^2}{\nu} = \left( \frac{d}{\eta_k}\right)^2\,,
  \label{eq:particle_reynolds}
\end{equation}
and, consequently, the aggregate is moving under creeping flow
conditions. The aggregate relaxation time is $\tau_r = (2
\varrho_p/\varrho_f + 1) d^2/(36\nu)$ and the Stokes number results
in:
\begin{equation}
  \Stk_p = \frac{\tau_r}{\tau_k} = \frac{(2 \varrho_p/\varrho_f + 1)}{36} \Rey_p
\,.  
\label{eq:particle_stokes}
\end{equation}
For solid-liquid dispersions, $\varrho_p/\varrho_f \sim 1$ and $\Stk_p
\ll 1$, the response of the particles to changes in
the flow field is very fast, confirming that aggregates inertia is negligible.

The previous assumptions are normally fulfilled in the turbulent
breakup of dilute solid-liquid colloidal dispersions. In this case
the size of the aggregates is less than few microns, whereas $\eta_k$
is seldom below 50 $\mu$m for mechanically stirred suspensions. Hence,
at every time these aggregates see a smooth and spatially well
correlated flow field around them. In such flow, the local turbulent
dissipation rate is related to the shear rate by the relationship:
\begin{equation}
  \varepsilon = \nu \dot\gamma^2\,.
  \label{eq:dissipation_shear}
\end{equation}
The shear rate is a function of the spatial velocity derivatives
through the rate of strain tensor $e^\infty_{ij}$:
\begin{equation}
  \dot\gamma = \sqrt{2 e^\infty_{ij} e^\infty_{ij}}\,, \qquad  \mbox{and} \qquad e^\infty_{ij} = \frac{1}{2}\left(\frac{\partial u_i}{\partial x_j}+\frac{\partial u_j}{\partial x_i} \right)\,.
  \label{eq:shear_rate}
\end{equation}
In fully developed turbulent flow, the local energy dissipation rate
is a highly fluctuating quantity. While the internal stresses acting
on the aggregate are determined by the interaction of the aggregate
structure with the flow field in its closest neighbourhood, the rate
of breakup depends also on the probability to enter a region of high
strain and hence on the motion of the aggregate over larger scales. To
capture both aspects the numerical investigation was performed at two
different levels. The fluid dynamics at the scales
comparable and larger than the Kolmogorov length scale and the
trajectories of the particles were calculated by a DNS of the
turbulent flow, in which the aggregates were modelled as point tracer
particles, due to their small size in comparison to the relevant
turbulent scales and their negligible inertia. Then, Stokesian
dynamics was adopted to evaluate the phenomena governed by the
smooth velocity field at the scales smaller than $\eta_k$. Thanks to
the information on the velocity gradients at the location of the
aggregates provided by the DNS and by taking into account the actual geometry of the
particles, Stokesian dynamics predicted both the rotational component of the motion of
the aggregates and the hydrodynamic forces acting on all
monomers. From this type of information, it is possible to evaluate
the internal forces and torques acting at intermonomer contacts and
determine the occurrence of breakup. As the aggregates studied in this
work are rigid and isostatic, rupture is caused by the failure of a
single intermonomer bond and we assumed that it occurs instantaneously
as soon as the tensile force on the most loaded bond exceeds the bond
strength.

\subsection{DNS of the turbulent flow}
\label{subsec:DNS}
For dilute suspensions of small particles, of size much smaller than
the Kolmogorov length and having negligible inertia, the geometrical
details of the particles can be smoothed out and they can be described
as point tracer particles passively carried by the turbulent flow. To
this aim, we used the results from the DNS of a statistically
stationary homogeneous and isotropic three-dimensional turbulent flow,
reported as Run II in the papers by \citet{Bec_2010a,Bec_2010b}. Data
are available from the iCFDdatabase (\verb#http://cfd.cineca.it#). We
used the trajectories of 3184 point tracer particles ($St_p=0$) that
were injected at random locations in the simulated turbulent
system. The equation of motion for inertialess point particles is
\begin{equation}
\dot{\bf X}(t)= {\bf u}({\bf X}(t),t)\,,
\label{eq:tracer}
\end{equation}
where ${\bf u}({\bf x},t)$ is the turbulent velocity field. Each
trajectory is sampled at 4720 different times: together with the
coordinates of the particle $X_i(t)$, the components of fluid velocity
$u_i({\bf X},t)$ and all the derivatives of fluid velocity at particle
position, $\partial u_i /\partial x_j$, are available.

The DNS solved the Navier-Stokes equations with large-scale
homogeneous and iso\-tropic forcing, on a cubic box with periodic
boundary conditions and $2048^3$ grid points. Extensive information on
the settings and methods used in the simulation is reported in the
cited papers by Bec and coworkers. The simulation assumed a very
dilute suspension, where particle-particle interactions and the
feedback of particles on the fluid flow were neglected.

\begin{table}
  \begin{center}
  \setlength{\tabcolsep}{8pt}
  \begin{tabular}{ccccccc}
    $\eta_k$, m         & $\tau_k$, s         & $\nu$, m$^2$/s       & $\left<\varepsilon\right>$, m$^2$s$^{-3}$ & 
    $\dot\gamma_{\rm eff}$, s$^{-1}$ & $\Delta t/\tau_k$    &  $\Rey_\lambda$\\
    $54.2\cdot 10^{-6}$ & $2.95\cdot 10^{-3}$ & $0.996\cdot 10^{-6}$ & 0.114                                     & 
    339                              & 0.05                 &  400 
   \end{tabular}
  \caption{Main parameters of the turbulent flow field simulated by
    DNS. $\eta_k$: Kolmogorov length scale; $\tau_k$: Kolmogorov time
    scale; $\nu$: kinematic viscosity; $\left<\varepsilon\right>$:
    average turbulent dissipation rate; $\dot\gamma_{\rm eff}$:
    effective mean shear rate; $\Delta t$: time interval at which each
    trajectory was sampled by DNS; $\Rey_\lambda$: Taylor-scale
    Reynolds number.}
  \label{tab:DNS_properties}
  \end{center}
\end{table}

The DNS variables are given in arbitrary computational units
\citep[see][table 1]{Bec_2010a} and were transformed to physical units
by prescribing values of $\nu$ and $\eta_k$ that are typical of
mechanically stirred aqueous suspensions. The main properties of the
turbulent field are listed in table~\ref{tab:DNS_properties}.

\begin{figure}%[!h]
  \centerline{\includegraphics[width=\textwidth]{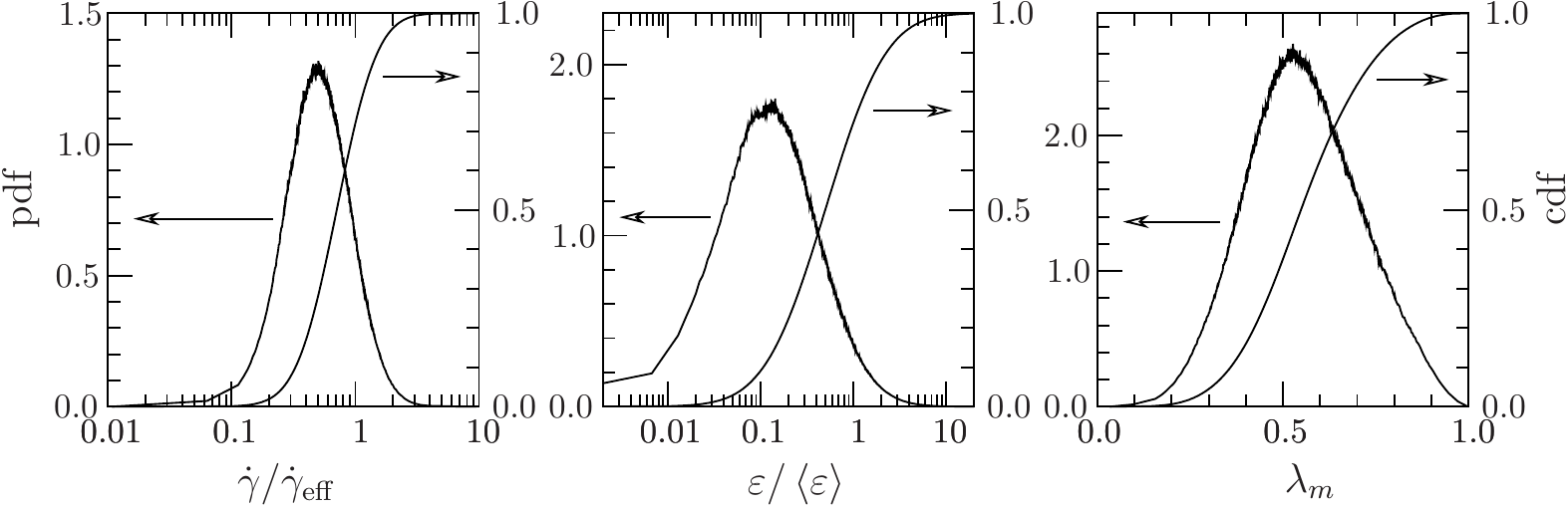}}
  \caption{Probability density functions (pdf values on the left axes)
    and cumulative distribution functions (cdf values on the right
    axes) of shear rate, $\dot\gamma$, turbulent dissipation rate,
    $\varepsilon$, and mixing index, $\lambda_m$, in the turbulent flow
    field.}
\label{fig:turb_properties}
\end{figure}

\begin{figure}%[!h]
  \centerline{\includegraphics[width=0.7\textwidth]{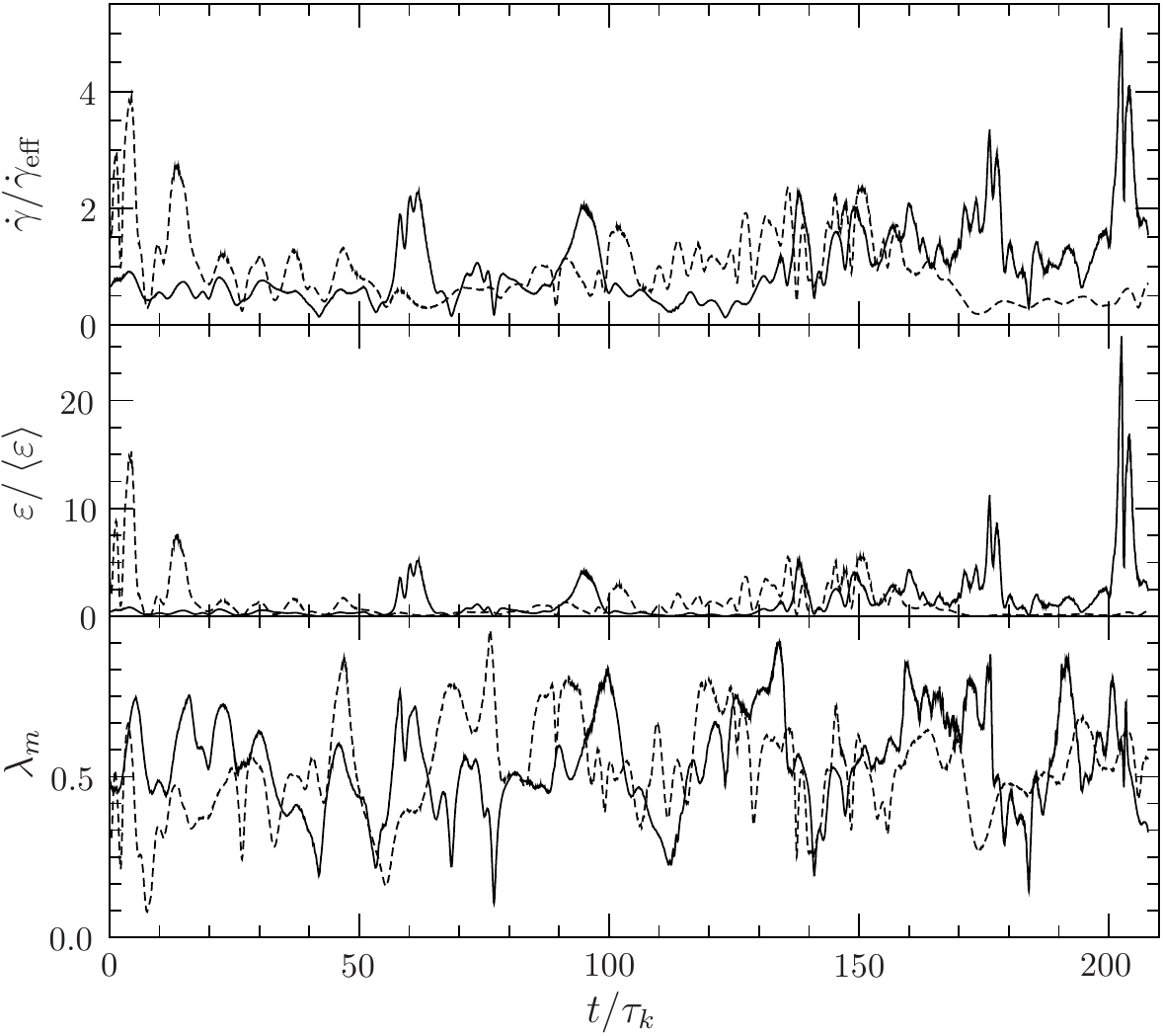}}
  \caption{Time series of shear rate, turbulence dissipation rate,
    mixing index along two different particle paths (continuous and
    dashed line, respectively).}
\label{fig:trajectories1}
\end{figure}

The analysis of the data along the trajectories made it possible to
reconstruct the features of the turbulence that are more relevant to
the breakup process and that are shown in
figure~\ref{fig:turb_properties}. The probability density functions
(pdf) of $\varepsilon$ and $\dot\gamma$, are approximately log-normal
\citep{Frisch_1995}. The log-normal shape of the pdf reflects the fact that
the spatial (and temporal) distributions of $\varepsilon$ or
$\dot\gamma$ are characterised by few sharp peaks of high intensity
between regions of low dissipation rate, as visible in
figure~\ref{fig:trajectories1}. Although it is customary to describe
turbulence in terms of $\varepsilon$, in this work we preferentially
use $\dot\gamma$, which is linked linearly to the hydrodynamic force
acting on the particles.

A practical approach to characterise the local flow type is provided
by the so-called mixing index $\lambda_m$ \citep{Manas-Zloczower_2009},
defined as follows:
\begin{equation}
  \lambda_m = \frac{\dot\gamma}{\dot{\gamma} + \tilde\omega} \,,
  \label{eq:mixing_index}
\end{equation}
where $\tilde\omega$ is the vorticity of the
fluid, that can be evaluated from the antisymmetrical component of the
velocity gradient:
\begin{equation}
  \tilde\omega = \sqrt{2 \omega^\infty_{ij} \omega^\infty_{ij}}\,,
  \qquad \mbox{and} \qquad \omega^\infty_{ij} =
  \frac{1}{2}\left(\frac{\partial u_i}{\partial x_j}-\frac{\partial
    u_j}{\partial x_i} \right)\,.
  \label{eq:rotation_rate}
\end{equation}
The mixing index $\lambda_m$ has a clear physical interpretation in
two-dimensional flows, which can be seen as the superposition of a
rotational and an elongational field, and $\lambda_m$ is the relative
weight of the elongational component of the flow. In three-dimensional
systems $\lambda_m$ has not such a direct interpretation, nevertheless
it still gives an indication of the elongational character of the
flow. In a turbulent flow, the mixing index varies from point to
point. The pdf of $\lambda_m$ is slightly skewed toward the
elongational region and, more importantly, shows that the probability of
highly rotational situations in the examined turbulence is very small,
whereas that of high elongation is more significant.

\subsection{Stokesian dynamics and internal stresses}
\label{subsec:SD}
The internal stresses in an aggregate are determined by the
interaction of the particle with the local flow field. As observed
before, in our case significant variations in the velocity gradient
take place on distances much larger than the size of the
aggregates. Consequently, in evaluating hydrodynamic stresses, it is
reasonable to neglect the curvature of the velocity profiles and
assume that the particles are surrounded by a linear flow field with
uniform velocity gradient. Under this assumption, Stokesian dynamics
\citep{Durlofsky_1987, Brady_1988} is an effective tool to determine
the hydrodynamic forces and torques acting on the aggregates. The
method is valid for conditions of Stokes flow and provides the
relationship between the hydrodynamic interactions acting on the
particles and their velocity. It is a meshless method, which
prescribes implicitly the condition of no slip at particle surfaces
and an undisturbed linear flow at infinity ${\mathbf u}^\infty({\mathbf x}) =
{\mathbf u}^\infty({\mathbf x}_{cm}) + {\mathbf \Gamma}^\infty\cdot
({\mathbf x}-{\mathbf x}_{cm})${ , where ${\mathbf x}_{cm}$ is the location of the centre of mass of the aggregate}. In our approach the aggregates 
translate as the point particles of the DNS and hence the
instantaneous value of the velocity gradient ${\mathbf \Gamma}^\infty$
in the fluid surrounding the aggregate is provided by the DNS of the flow field.

By using Stokesian dynamics the hydrodynamic force, torque and
stresslet acting on monomer $i$ of an aggregate made of $p$ spherical
monomers can be expressed as a linear combination of the linear and
rotational velocities of all the monomers relative to the imposed flow
and the rate of strain:
{ 
\begin{multline}
\label{eq:StokDyn1}
\left[
%      \begin{array}{ccc|ccc|ccc}
      \begin{array}{ccc ccc ccc}
        {\mathbf m}_{11}^{uf} & 
        \cdots &
        {\mathbf m}_{1p}^{uf} &
        {\mathbf m}_{11}^{ul} & 
        \cdots &
        {\mathbf m}_{1p}^{ul} &
        {\mathbf m}_{11}^{us} & 
        \cdots &
        {\mathbf m}_{1p}^{us} \\
        \cdots &
        \cdots &
        \cdots &
        \cdots &
        \cdots &
        \cdots &
        \cdots &
        \cdots &
        \cdots \\
        {\mathbf m}_{p1}^{uf} & 
        \cdots &
        {\mathbf m}_{pp}^{uf} &
        {\mathbf m}_{p1}^{ul} & 
        \cdots &
        {\mathbf m}_{pp}^{ul} &
        {\mathbf m}_{p1}^{us} & 
        \cdots &
        {\mathbf m}_{pp}^{us} \\
%        \hline 
        {\mathbf m}_{11}^{\omega f} & 
        \cdots &
        {\mathbf m}_{1p}^{\omega f} &
        {\mathbf m}_{11}^{\omega l} & 
        \cdots &
        {\mathbf m}_{1p}^{\omega l} &
        {\mathbf m}_{11}^{\omega s} & 
        \cdots &
        {\mathbf m}_{1p}^{\omega s} \\
        \cdots &
        \cdots &
        \cdots &
        \cdots &
        \cdots &
        \cdots &
        \cdots &
        \cdots &
        \cdots \\
        {\mathbf m}_{p1}^{\omega f} & 
        \cdots &
        {\mathbf m}_{pp}^{\omega f} &
        {\mathbf m}_{p1}^{\omega l} & 
        \cdots &
        {\mathbf m}_{pp}^{\omega l} &
        {\mathbf m}_{p1}^{\omega s} & 
        \cdots &
        {\mathbf m}_{pp}^{\omega s} \\
%        \hline
        {\mathbf m}_{11}^{ef} & 
        \cdots &
        {\mathbf m}_{1p}^{ef} &
        {\mathbf m}_{11}^{el} & 
        \cdots &
        {\mathbf m}_{1p}^{el} &
        {\mathbf m}_{11}^{es} & 
        \cdots &
        {\mathbf m}_{1p}^{es} \\
        \cdots &
        \cdots &
        \cdots &
        \cdots &
        \cdots &
        \cdots &
        \cdots &
        \cdots &
        \cdots \\
        {\mathbf m}_{p1}^{ef} & 
        \cdots &
        {\mathbf m}_{pp}^{ef} &
        {\mathbf m}_{p1}^{el} & 
        \cdots &
        {\mathbf m}_{pp}^{el} &
        {\mathbf m}_{p1}^{es} & 
        \cdots &
        {\mathbf m}_{pp}^{es}
      \end{array}
\right] 
\cdot \\ 
\left\{
      \begin{array}{c}
        {\mathbf F}_1 \\
        \cdots \\
        {\mathbf F}_{p} \\
%        \hline
        {\mathbf L}_1 \\
        \cdots \\
        {\mathbf L}_{p} \\
%        \hline
        {\mathbf S}_1 \\
        \cdots \\
        {\mathbf S}_{p} 
      \end{array}
\right\}
= -
\left\{
      \begin{array}{c}
        {\mathbf u}_1 - {\mathbf u}^\infty({\mathbf x}_1) \\
        \cdots \\
        {\mathbf u}_{p} - {\mathbf u}^\infty({\mathbf x}_{p}) \\
%        \hline
        {\boldsymbol \omega}_1 - {\boldsymbol \omega}^\infty({\mathbf x}_1)\\
        \cdots \\
        {\boldsymbol \omega}_{p} - {\boldsymbol \omega}^\infty({\mathbf x}_{p}) \\
%        \hline
        - {\mathbf e}^\infty({\mathbf x}_1) \\
        \cdots \\
        - {\mathbf e}^\infty({\mathbf x}_{p}) 
      \end{array}
\right\}
\end{multline}
}

In the equation above, the entries ${\mathbf m}_{ij}^{kq}$, with
$i,j=1,\dots,p$ and $k,q=u,\omega,e,f,l,s$ form the so-called mobility
matrix coupling particle velocities to hydrodynamics forces. Moreover,
${\mathbf u}_i$ and ${\boldsymbol \omega}_i$ are the linear and
angular velocities of the primary particle $i$, ${\mathbf F}_i$,
${\mathbf L}_i$ and ${\mathbf S}_i$ the hydrodynamic force, torque and
stresslet acting on the primary particle; the linear and angular
velocities and the deformation rate of the undisturbed flow, ${\mathbf
  u}^\infty$, ${\boldsymbol \omega}^\infty$, ${\mathbf e}^\infty$,
must be evaluated at the centre ${\mathbf x}_i$ of the
monomer. Forces, torques and velocities entering
eq.~(\ref{eq:StokDyn1}) are three-component column vectors, while rate
of strain and stresslet tensors in the right hand side of the equation
are conveniently represented as five-component vectors by taking
advantage of the fact that they are symmetric and traceless:
\begin{equation}
\label{eq:strain_stresslet}
{\mathbf e}^\infty = 
\left\{
      \begin{array}{c}
        e^\infty_{xx}-e^\infty_{zz} \\
        2e^\infty_{xy} \\
        2e^\infty_{xz} \\
        2e^\infty_{yz} \\
        e^\infty_{yy}-e^\infty_{zz}
      \end{array}
\right\}\,,   
\qquad
{\mathbf S}_i = 
\left\{
      \begin{array}{c}
        S_{i,xx} \\
        S_{i,xy} \\
        S_{i,xz} \\
        S_{i,yz} \\
        S_{i,yy} 
      \end{array}
\right\}
.   
\end{equation}
In this way, all redundant components are removed from the algebraic
system of equations and the simulation of a cluster of $p$ primary
particles leads to a $11 p \times 11 p$ mobility matrix. This is
symmetric, positive definite and depends only on the size and
relative position of the monomers and linearly on viscosity
\citep{Durlofsky_1987}.

When Stokesian dynamics is applied to suspensions of freely 
moving particles, the mobility matrix is normally generated by 
combining a low order multipole expansion of the Stokes solution, 
capable of capturing accurately the hydrodynamic interaction
when the particles are relatively far apart, and a correction for near-field 
interaction, based on lubrication theory. In our approach the latter term was 
neglected, because lubrication has an effect only on approaching 
particles, whereas in our rigid clusters all the monomers move at 
the same velocity, and the addition of the lubrication 
terms can lead to significant overestimation of the friction force 
\citep{Bossis_1991}. Therefore, we used the far-field form of the mobility matrix, which is obtained by pairwise additivity of the far-field mobility of individual particles. The expressions of the terms of this matrix as functions of the geometric configuration are reported in the literature \citep{Durlofsky_1987,Ichiki_2008}. {  In spite of the approximation involved, the method based on the far-field mobility matrix gives accurate predictions of the motion of rigid clusters, as shown by calculations of the settling velocity of compact aggregates \citep[Supplementary Material]{Harsche_2010} and the rotational velocity of rigid chains of spheres in a shear flow \citep[Supplementary Material]{Vanni_2011}.} 

The velocity of each primary particle $i$ can be related to the linear
and angular velocity of the centre of mass ($cm$) of the cluster by
the condition of rigid body motion:
\begin{equation}
\label{eq:RigidBodyCondition}
\left\{
      \begin{array}{c c l}
            \mathbf{u}_i      & = & \mathbf{u}_{cm}   + \boldsymbol{\omega}_{cm} \times (\mathbf{x}_i-\mathbf{x}_{cm}) \\
            \boldsymbol{\omega}_i & = & \boldsymbol{\omega}_{cm} 
      \end{array}
\right.  \qquad\qquad i=1,2, \ldots, p \, .
\end{equation}
In addition, in an inertialess aggregate all forces and torques must be balanced:
\begin{equation}
\label{eq:ForceBalance}
\left\{
      \begin{array}{l}
            \sum_{i=1}^{p} \mathbf{F}_i  = 0 \\
            \sum_{i=1}^{p} \left[\mathbf{L}_i + (\mathbf{x}_i-\mathbf{x}_{cm})\times \mathbf{F}_i\right] = 0
      \end{array} 
\right.
\end{equation}

By substituting Eq.~(\ref{eq:RigidBodyCondition}) in
Eq.~(\ref{eq:StokDyn1}) and adding Eq.~(\ref{eq:ForceBalance}) to the
resulting system, one finally obtains the following linear system,
whose solution gives forces, torques and stresslets acting on the
individual monomers, and the translational and angular velocity of the
aggregate:
{ 
\begin{multline}
\label{eq:StokDyn2}
\left[
%      \begin{array}{ccc|ccc|ccc|cc}
      \begin{array}{ccc ccc ccc cc}
      {\mathbf m}_{11}^{uf} & 
        \cdots &
        {\mathbf m}_{1p}^{uf} &
        {\mathbf m}_{11}^{ul} & 
        \cdots &
        {\mathbf m}_{1p}^{ul} &
        {\mathbf m}_{11}^{us} & 
        \cdots &
        {\mathbf m}_{1p}^{us} &
        {\mathbf I} &
        {\boldsymbol \alpha}_1 \\
        \cdots &
        \cdots &
        \cdots &
        \cdots &
        \cdots &
        \cdots &
        \cdots &
        \cdots &
        \cdots &
        \cdots &
        \cdots \\
        {\mathbf m}_{p1}^{uf} & 
        \cdots &
        {\mathbf m}_{pp}^{uf} &
        {\mathbf m}_{p1}^{ul} & 
        \cdots &
        {\mathbf m}_{pp}^{ul} &
        {\mathbf m}_{p1}^{us} & 
        \cdots &
        {\mathbf m}_{pp}^{us} &
        {\mathbf I} &
        {\boldsymbol \alpha}_{p} \\
%     \hline
        {\mathbf m}_{11}^{\omega f} & 
        \cdots &
        {\mathbf m}_{1p}^{\omega f} &
        {\mathbf m}_{11}^{\omega l} & 
        \cdots &
        {\mathbf m}_{1p}^{\omega l} &
        {\mathbf m}_{11}^{\omega s} & 
        \cdots &
        {\mathbf m}_{1p}^{\omega s} &
        {\mathbf 0} &
        {\mathbf I} \\
        \cdots &
        \cdots &
        \cdots &
        \cdots &
        \cdots &
        \cdots &
        \cdots &
        \cdots &
        \cdots &
        \cdots &
        \cdots \\
        {\mathbf m}_{p1}^{\omega f} & 
        \cdots &
        {\mathbf m}_{pp}^{\omega f} &
        {\mathbf m}_{p1}^{\omega l} & 
        \cdots &
        {\mathbf m}_{pp}^{\omega l} &
        {\mathbf m}_{p1}^{\omega s} & 
        \cdots &
        {\mathbf m}_{pp}^{\omega s} &
        {\mathbf 0} &
        {\mathbf I} \\
%      \hline
        {\mathbf m}_{11}^{ef} & 
        \cdots &
        {\mathbf m}_{1p}^{ef} &
        {\mathbf m}_{11}^{el} & 
        \cdots &
        {\mathbf m}_{1p}^{el} &
        {\mathbf m}_{11}^{es} & 
        \cdots &
        {\mathbf m}_{1p}^{es} &
        {\mathbf 0} &
        {\mathbf 0} \\
        \cdots &
        \cdots &
        \cdots &
        \cdots &
        \cdots &
        \cdots &
        \cdots &
        \cdots &
        \cdots &
        \cdots &
        \cdots \\
        {\mathbf m}_{p1}^{ef} & 
        \cdots &
        {\mathbf m}_{pp}^{ef} &
        {\mathbf m}_{p1}^{el} & 
        \cdots &
        {\mathbf m}_{pp}^{el} &
        {\mathbf m}_{p1}^{es} & 
        \cdots &
        {\mathbf m}_{pp}^{es} &
        {\mathbf 0} &
        {\mathbf 0} \\
%      \hline
        {\mathbf I} &
        \cdots &
        {\mathbf I} &
        {\mathbf 0} &
        \cdots &
        {\mathbf 0} &
        {\mathbf 0} &
        \cdots &
        {\mathbf 0} &
        {\mathbf 0} &
        {\mathbf 0} \\
        {\boldsymbol \alpha}_1^T &
        \cdots &
        {\boldsymbol \alpha}_{p}^T &
        {\mathbf I} &
        \cdots &
        {\mathbf I} &
        {\mathbf 0} &
        \cdots &
        {\mathbf 0} &
        {\mathbf 0} &
        {\mathbf 0} 
      \end{array}
\right]
\cdot \\
\left\{
      \begin{array}{c}
        {\mathbf F}_1 \\
        \cdots \\
        {\mathbf F}_{p} \\
%        \hline
        {\mathbf L}_1 \\
        \cdots \\
        {\mathbf L}_{p} \\
%        \hline
        {\mathbf S}_1 \\
        \cdots \\
        {\mathbf S}_{p} \\
%        \hline
        {\mathbf u}_{cm} \\
        {\boldsymbol \omega}_{cm} 
      \end{array}
\right\}
= 
\left\{
      \begin{array}{c}
        {\mathbf u}^\infty({\mathbf x}_1) \\
        \cdots \\
        {\mathbf u}^\infty({\mathbf x}_{p}) \\
%        \hline
        {\boldsymbol \omega}^\infty({\mathbf x}_1)\\
        \cdots \\
        {\boldsymbol \omega}^\infty({\mathbf x}_{p}) \\
%        \hline
        {\mathbf e}^\infty({\mathbf x}_1) \\
        \cdots \\
        {\mathbf e}^\infty({\mathbf x}_{p}) \\
%        \hline 
        {\mathbf 0} \\
        {\mathbf 0} 
      \end{array}
\right\}
\end{multline}
}

In Eq.~(\ref{eq:StokDyn2}) $\mathbf{0}$ and $\mathbf{I}$ are the $3 \times 3$ null and identity matrices, respectively, and 
$$
\boldsymbol{\alpha}_i = \left[
      \begin{array}{ccc}
                 0     &   (z_i-z_{cm}) & - (y_i-y_{cm}) \\
        - (z_i-z_{cm}) &          0     &   (x_i-x_{cm}) \\
          (y_i-y_{cm}) & - (x_i-x_{cm}) &          0     
      \end{array}
                     \right]
$$

The linear system of Eq.~(\ref{eq:StokDyn2}) has $11 p + 6$ unknowns
and its matrix is symmetric but no longer positive definite. It was
solved at every time step by using the LAPACK routines \textsf{dsytrf}
and \textsf{dsytrs} for factorization and backsubstitution, 
respectively \citep{Anderson_1999}. It is convenient to
consider the equations in a reference frame moving with the cluster
(quasi-Lagrangian frame), because then the matrix does not
change with the orientation of the aggregate and thus only a single
factorization is needed.

An isostatic aggregate made by $p$ primary particles shows $p-1$ 
interparticle contacts and the internal
stresses at these contacts can be evaluated simply by solving force
and moment balances on each primary particle, without the need to
understand in detail the nature of the inter-monomer interaction
\citep{Gastaldi_2011, Seto_2011}. For the $i-$th monomer of the
aggregate, these balances read as follows:
\begin{equation}
 {\mathbf F}_i  + \sum_{j=1}^{p} \beta_{j,i}{\mathbf f}_{j,i} = 0 \qquad\qquad i=1,\ldots,p
 \label{eq:bforces}
\end{equation}
\begin{equation}
{\mathbf L}_i  + \sum_{j=1}^{p} \beta_{j,i} \left[ {\mathbf l}_{j,i}  +  \frac{{\mathbf x}_j-{\mathbf x}_i}{2} \times{\bf f}_{j,i} \right]= 0 \qquad\qquad i=1,\ldots,p 
\label{eq:btorques}
\end{equation}
where ${\mathbf f}_{j,i}$ and ${\mathbf l}_{j,i}$ are the internal
force and torque acting on monomer $i$ at the contact with monomer $j$
and $\beta_{j,i}$ is a variable which is set equal to 1 if monomers
$i$ and $j$ are in contact and 0 otherwise. Only $6(p-1)$ out of the
$6 p$ equations above are independent, because the balance of force
and moment for the whole aggregate given by
eq.~(\ref{eq:ForceBalance}) prescribes that the sums of all internal
forces and of all internal moments must vanish. Therefore the solution
of the linear system obtained by taking the equations above for the
first $p-1$ particles gives all the required components of the
internal forces and torques acting on the $p-1$ internal bonds. The
force acting at each contact is then decomposed into a normal
component $N$, acting on the line joining the centres of the monomers,
and a tangential component $T$, acting in the perpendicular
direction. Similarly, the torque is decomposed in a twisting moment
$M_t$, which generates the spinning deformation around the
centre-centre axis, and a bending moment $M_b$, which induces mutual
rolling of the two monomers.

The translational and rotational aggregate velocities should be obtained by solving the
system of equations~(\ref{eq:StokDyn2}), which require the knowledge of the gradient velocity 
tensor fields. Here we make the hypothesis that the center of mass translates with 
the flow velocity at its position:
\begin{equation}
\mathbf{u}_{cm}(t) = \mathbf{u}(\mathbf{x}_{cm}(t),t)\,.
\label{eq:masscentervel}
\end{equation}
This assumption does not give rise to any error if the particle is axisymmetric, as in the 
case of doublets, but is not fully correct in the case of disordered aggregates, whose
velocity may be different from that of fluid particles. However, the error is minor and does 
not invalidate the results. For example, with the clusters of 384 primary particles 
studied in section~\ref{sec:clu-clu}, the estimated error 
in particle position is always less than $d/10$ ($d$ being the outer diameter 
of the aggregate). As we assumed $d \ll \eta_k$, such a displacement is much smaller 
than the Kolmogorov length scale and consequently the cluster is always located in 
a region where the velocity gradient does not deviate significantly from the value 
of the DNS trajectory. Hence, the translational motion of the aggregate was
provided by the DNS. Differently, the rotation around its centre of
mass was calculated by integration of the angular velocity $\boldsymbol{\omega}_{cm}$ 
obtained by eq~(\ref{eq:StokDyn2})
by an explicit Euler method with a time step $10$ times smaller than the
sampling time of DNS trajectories. At every time step, the angular
motion consisted of a rotation of the angle $\Delta \varphi =
\left|\boldsymbol{\omega}_{cm}(t)\right| \Delta t $ around the unit
vector $\mathbf{n} =
\boldsymbol{\omega}_{cm}/\left|\boldsymbol{\omega}_{cm}\right|$. 
The new coordinates of the monomers after the
operation are given by the following equation \citep{Goldstein_1983}:
\begin{equation} 
\label{eq:rotation} 
\mathbf{x}_{i}(t+\Delta t) = \mathbf{x}_{cm}(t+\Delta t) + \mathbf{\cal R}\cdot \left(\mathbf{x}_i(t) - \mathbf{x}_{cm}(t) \right) \, ,
\end{equation} 
where the rotation matrix is: 
\begin{equation} 
\begin{scriptsize} 
\mathbf{\cal R} = 
\left[ 
      \begin{array}{ccc} 
        \cos\Delta\varphi + n_{x}^2 (1-\cos\Delta\varphi) & 
        n_{x} n_{y}(1-\cos\Delta\varphi) - n_{z}\sin\Delta\varphi & 
        n_{x} n_{z}(1-\cos\Delta\varphi) + n_{y}\sin\Delta\varphi \\ 
        n_{y} n_{x}(1-\cos\Delta\varphi) + n_{z}\sin\Delta\varphi & 
        \cos\Delta\varphi + n_{y}^2 (1-\cos\Delta\varphi) & 
        n_{y} n_{z}(1-\cos\Delta\varphi) - n_{x}\sin\Delta\varphi \\ 
        n_{z} n_{x}(1-\cos\Delta\varphi) - n_{y}\sin\Delta\varphi & 
        n_{z} n_{y}(1-\cos\Delta\varphi) + n_{x}\sin\Delta\varphi & 
        \cos\Delta\varphi + n_{z}^2 (1-\cos\Delta\varphi)  
      \end{array} 
\right] 
\end{scriptsize} 
\end{equation} 

\subsection{Breakup frequency} \label{sec:break_frequency}
Since the studied aggregates are not statically overconstrained, their
breakage is originated by the failure of a single bond and the
subsequent separation of the two contacting monomers, giving two
fragments. The failure occurs whenever the normal force acting on the
bond is tensile and exceeds the critical pull-off value given by
contact mechanics \citep{Johnson_1985}:
\begin{equation}
 N_{cr} = \kappa \pi a \sigma \,,
\label{eq:pulloff}
\end{equation}
where $\sigma$ is the surface energy of the contact, $a$ the radius of
the contacting primary particles, and $\kappa$ is a parameter that
ranges between 1.5 and 2.0 depending on the physical properties of the
contact \citep{Maugis_1992,Carpick_1999}. As reviewed by
\citet{Baebler_2008a}, a number of experimental and theoretical works
suggests that rigid aggregates are brittle and hence breakup occurs
almost instantaneously as soon as the pull off force is exceeded.

In this work we apply first-passage time statistics
\citep{Redner_2001} to estimate the breakup frequency from the
distribution of the time necessary for observing the first occurrence
of a normal stress strong enough to severe a bond and, consequently,
break the aggregate. Considering an aggregate injected in the flow
field at random space and time coordinates, its breakup time $\tau$ is the
time needed for it to reach the condition $N>N_{cr}$ on any of its
bonds for the first time (provided that at injection time
$N<N_{cr}$). The breakup frequency for a class of aggregates is the
inverse of the mean breakup time, obtained by averaging $\tau$ over many
different injections and aggregate realisations \citep{Baebler_2012}:
\begin{equation}
 f_{br} = \frac{1}{\left< \tau \right>}\,.
 \label{eq:f_br}
\end{equation}

\begin{figure}%[!h]
  \centerline{\includegraphics[width=0.7\textwidth]{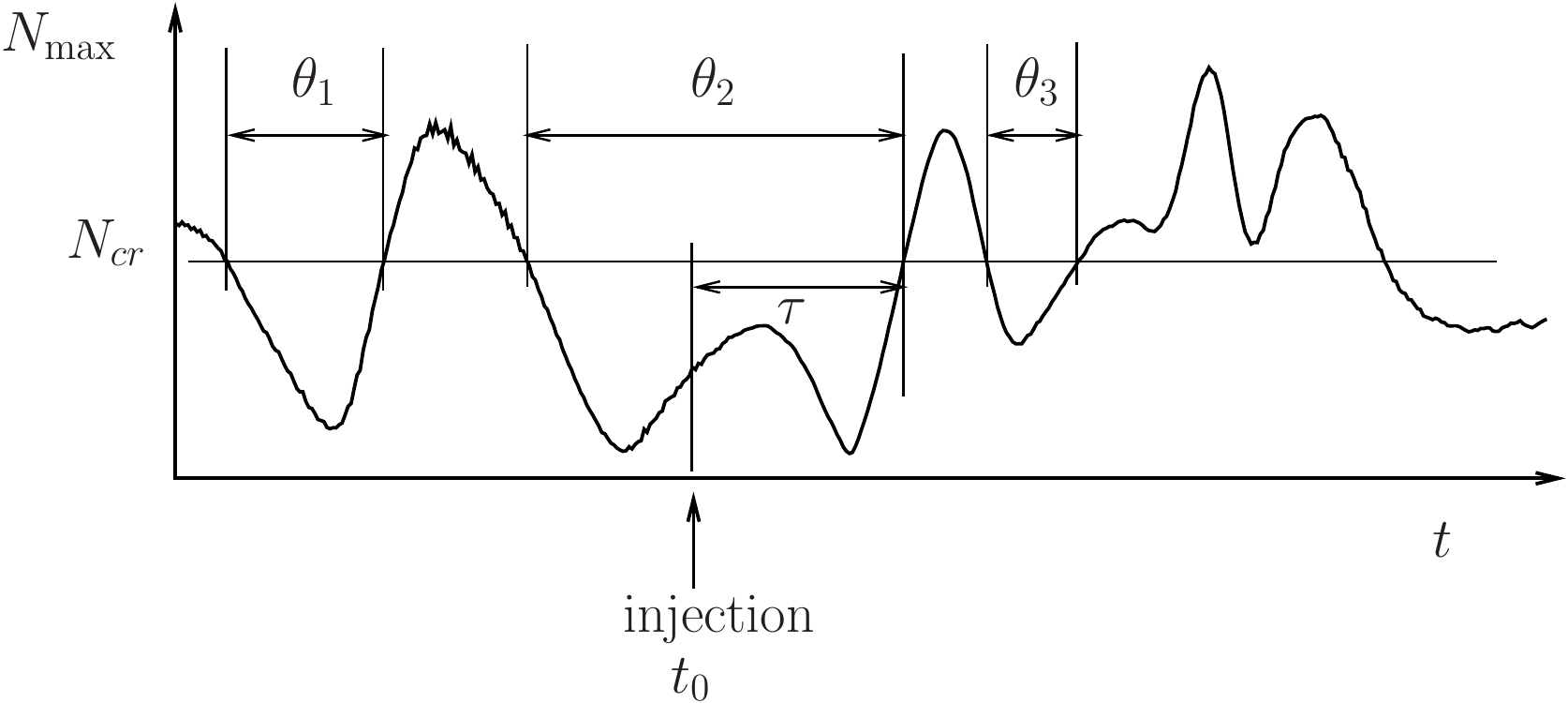}}
  \caption{Schematic of the time series of the instantaneous maximum
    normal stress $N_{\rm max}$ acting on an aggregate that moves in
    the turbulent flow field. If the aggregate were injected in the
    field at time $t_0$ its breakage time would be $\tau$. The diving
    times $\theta_1$, $\theta_2$, $\theta_3$ for a given stress
    threshold $N_{cr}$ are also shown.}
\label{fig:time_series}
\end{figure}

The mean breakup time can be obtained by the analysis of the time
series of the largest instantaneous normal stress acting on each
aggregate, $N_{\rm max}$. An example of such a time series is plotted
in figure~\ref{fig:time_series}. Given the pull-off threshold $N_{cr}$
for the normal stress, one can measure the sequence of {\it diving times}
$\theta_1$, $\theta_2$, $\theta_3$, \dots, that is, the lengths of the
intervals of time in which $N_{\rm max}$ is below the threshold
$N_{cr}$ and breakup does not happen. The moments of the distribution
of diving times for all aggregates can be related to the mean breakage
time as follows \citep{Baebler_2012}:
\begin{equation}
\left< \tau \right> = \frac{\left< \theta^2(N_{cr}) \right>}{2 \left< \theta(N_{cr}) \right>}\,.
 \label{eq:breakage_time}
\end{equation}
Finally, by sampling the size of the two fragments that are formed
each time an aggregate is broken (that is, whenever $N_{\rm max}$
becomes greater than $N_{cr}$), one can estimate the fragment
distribution function of the process.

The approach is similar to the one adopted recently by
\citet{Baebler_2012}. In that case, however, it was assumed that
breakup occurred whenever a critical value of dissipation rate was
overcome and the role of internal stresses was not taken into
account. Here instead, breakup is determined by the combined effect of
the dissipation rate and the orientation of the aggregate, which
determine the internal stress. As shown below, the interaction between
the two effects can be relevant and should be taken into account. In
addition, by considering the structure of the aggregate it is possible
to identify the failed contact and to extract statistics on the
fragment size distribution.

The non-normal interactions are not capable of breaking the aggregate
but may lead to its restructuring. This process may proceed through
mutual sliding of the contacting monomers induced by transverse
forces, relative twisting by torsional moments or mutual rolling due
to bending moments. For aggregates made of spherical monomers, the
most relevant restructuring mechanism is mutual rolling, which is
determined by a critical value in the bending moment
\citep{Vanni_2011}. However, the theory to determine the onset of
rolling is not so well established as the method to predict particle
detachment. Therefore, although in this paper we give indications on the magnitude
and the distribution of the bending moment, the process of
restructuring is not examined in detail.

\section{A simple configuration: the doublet}
\label{sec:doublet}
As a first step we investigated the response of a rigid doublet of equal
sized and contacting spheres to turbulence. In the study of breakup,
the doublet has been used several times as the prototype of aggregates
that break in two fragments because of the presence of a weak region
in between
\citep{Calvert_2009,Dukhin_2005,Manas-Zloczower_2009,Tadmor_1976}. Furthermore,
the problem of the motion of a doublet in linear velocity fields has
an exact analytical solution and thus allows validation of our method
based on Stokesian dynamics.

\subsection{Validation of the method}

The rotational motion of an axisymmetric body in a three-dimensional linear flow field at creeping
flow conditions obeys the following equation
\citep[p. 75]{Guazzelli_2012}:
\begin{equation}
\label{eq:axi_rotation}
\frac{{\rm d} \mathbf p}{{\rm d}t} = {\boldsymbol \omega}^\infty
     {\boldsymbol \times} {\mathbf p} + \beta \left[ {\mathbf
         e}^\infty {\boldsymbol \cdot} {\mathbf p} - {\mathbf p}
       \left( {\mathbf p}{\boldsymbol \cdot}{\mathbf e}^\infty
            {\boldsymbol \cdot}{\mathbf p}\right) \right]\,,
\end{equation}
where the director $ \mathbf p$ is the unit vector in the direction of
the symmetry axis and the Bretherton constant \citep{Bretherton_1962}
for a doublet made of equal sized and contacting spheres is $\beta=0.5942$.

When applied to a doublet positioned on the $x-y$ plane and immersed
in a simple shear flow with prescribed velocity gradient $\dot{\gamma}
= \frac{{\rm d} u^\infty_x}{{\rm d}y}$, the previous equation gives
the following expression for the angular velocity of the body:
\begin{equation}
\label{eq:omega_nir}
\omega_x = \omega_y = 0 \qquad \omega_z =
-\frac{1}{2}\dot{\gamma}\left[1+\beta\cos(2\phi)\right]\,,
\end{equation}
where the angle $\phi$ defines the orientation of the doublet in the
plane $x-y$ in a fixed frame of reference.

The internal bending and twisting moments acting at the contact region
of the two monomers, due to symmetry, are always nil for a freely
suspended doublet. Instead the normal and tangential components of the
hydrodynamic force acting on either sphere of the doublet are given by
the following expressions, obtained by applying the solution for a general linear flow by \citet{Nir_1973} to our configuration:
\begin{eqnarray}
\label{eq:forceN_nir}
N &=& \pi\mu\frac{a^2}{2}\dot{\gamma}\left(h_1+h_2\right)\sin(2\phi)\,,\\
\label{eq:forceT_nir}
T &=& \pi\mu\frac{a^2}{2}\dot{\gamma} h_1 \cos(2\phi)\,,
\end{eqnarray}
where $h_1 = 4.463$, $h_2 = 7.767$. Therefore the
doublet simply rotates in the $x-y$ plane and alternates intervals of
traction and compression. {  As the doublet is rigid, the two monomers always remain in contact and cannot move apart during traction}.

\begin{figure}
  \centerline{\includegraphics[width=0.9\textwidth]{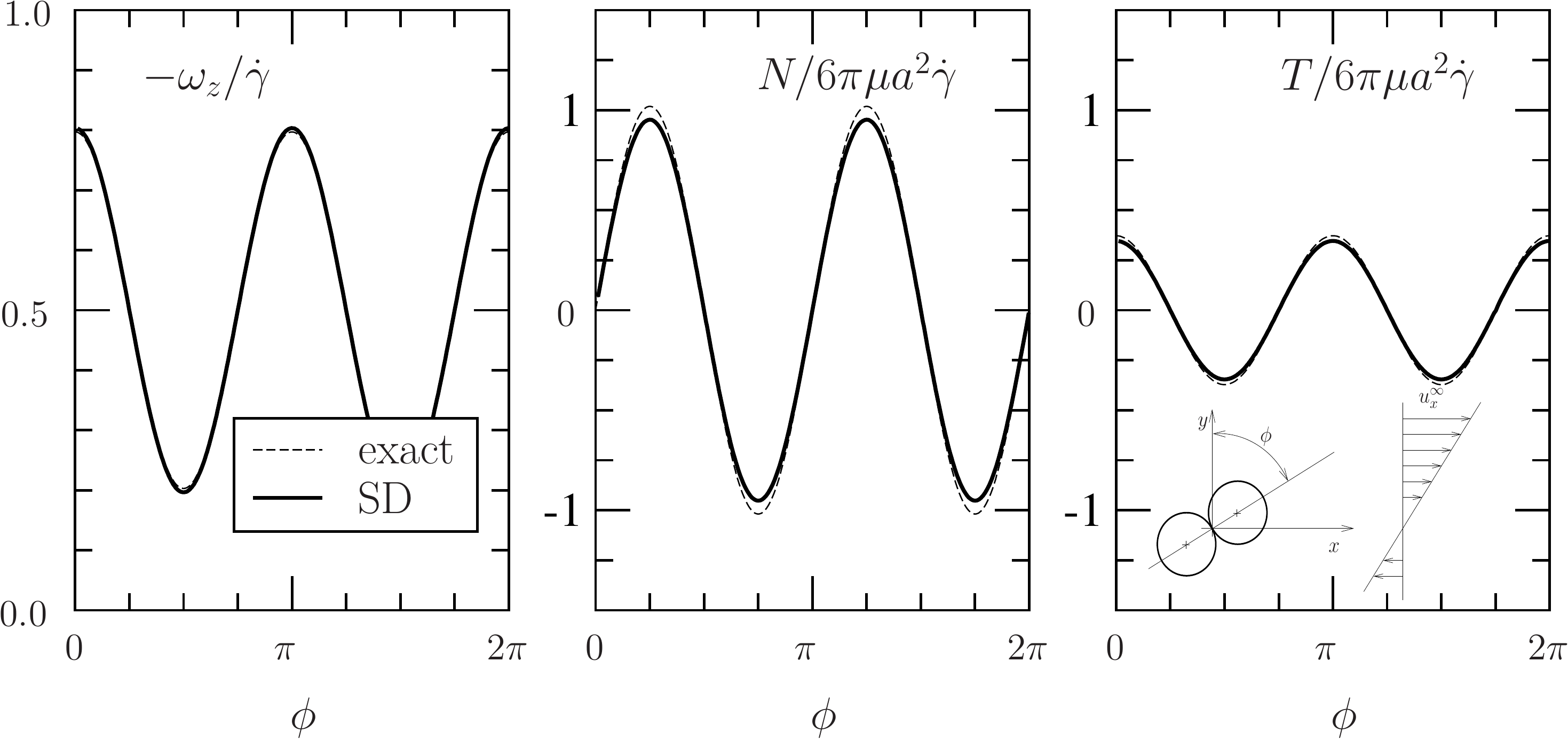}}
  \caption{Angular velocity and hydrodynamic forces for a doublet
    versus orientation. Dashed line: exact solutions,
    eqs~(\ref{eq:omega_nir}), (\ref{eq:forceN_nir}),
    (\ref{eq:forceT_nir}); continuous line: Stokesian dynamics.}
\label{fig:dimeroFTS}
\end{figure}

As apparent from figure~\ref{fig:dimeroFTS}, the Stokesian dynamics data
agree well with the exact solution for the rotational velocity of the
doublet with an error around 1\%. The estimated dimensionless doublet
rotation period is $t_r\dot{\gamma} \, = 15.8$, against the
theoretical value $t_r\dot{\gamma} \, = 2 \pi (q+1/q) = 15.6$. The
variable $q$ is the eccentricity of the ellipsoid that moves with the
same angular velocity as our doublet and is related to the Bretherton
constant by: $\beta= (q^2-1)/(q^2+1)$. Concerning the internal forces
$N$ and $T$, the difference between Stokesian dynamics and exact
solution is larger (about 6\%), but still fairly good.

\begin{figure}%[!h]
  \centerline{\includegraphics[width=0.8\textwidth]{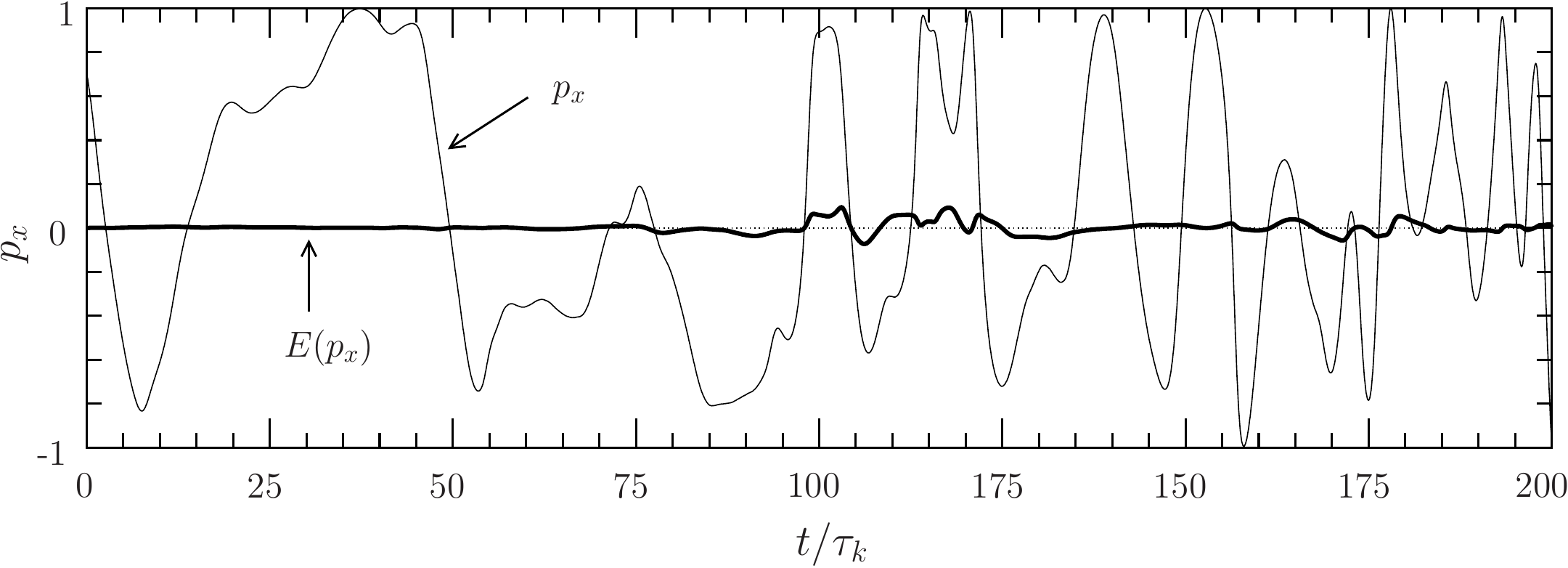}}
  \caption{$x-$component of the director vector of the doublet, $p_x$, for one
    turbulent trajectory calculated by accurate solution of
    eq.~(\ref{eq:axi_rotation}) and absolute error of the Stokesian
    dynamics simulation, $E(p_x)$.}
\label{fig:error_turb}
\end{figure}

For any given linear flow field, the instantaneous values of angular
velocity and internal forces depend only on the orientation of the
doublet with respect to the velocity gradient. As a consequence, it is
important to characterise the error in the orientation and its
accumulation in time during the motion of the particle in the
turbulent flow. This is why we compared the orientation of the doublet
given by the Stokesian dynamics method with the result of an accurate
integration of the ordinary differential equation
(\ref{eq:axi_rotation}) for some of the particle paths provided by the
DNS. A result showing the behaviour of the $x$-component of the
director vector is reported in figure~\ref{fig:error_turb}. The error
of Stokesian dynamics with respect to the accurate integration is
always below 4-5\%, even in the presence of sudden changes of
orientation. { The component of such error due to the numerical
  discretisation is almost negligible (about 3\% of the total error),
  as verified by integrating the equations with much smaller time
  steps. In addition no significant degradation of the rotation matrix
  $\mathbf{\cal R}$ was found at the end of the integration. Hence,
  the error is essentially generated by the approximate mobility
  matrix, which is estimated from a multipole expansion of the
  solution of the Stokes equations truncated to first order. As
  observed before for a doublet aligned in the shear plane, Stokesian
  dynamics gives errors around 1 \% in the prediction of the angular
  velocity.} However, the rotation of the particle and the fluctuating
flow field act in such a way as to balance the error and avoid its
increase as the simulation time becomes larger.

\subsection{Stress distribution and breakup rate}
\label{subsec:stressdoublet}
\begin{figure}%[!h]
  \centerline{\includegraphics[width=0.7\textwidth]{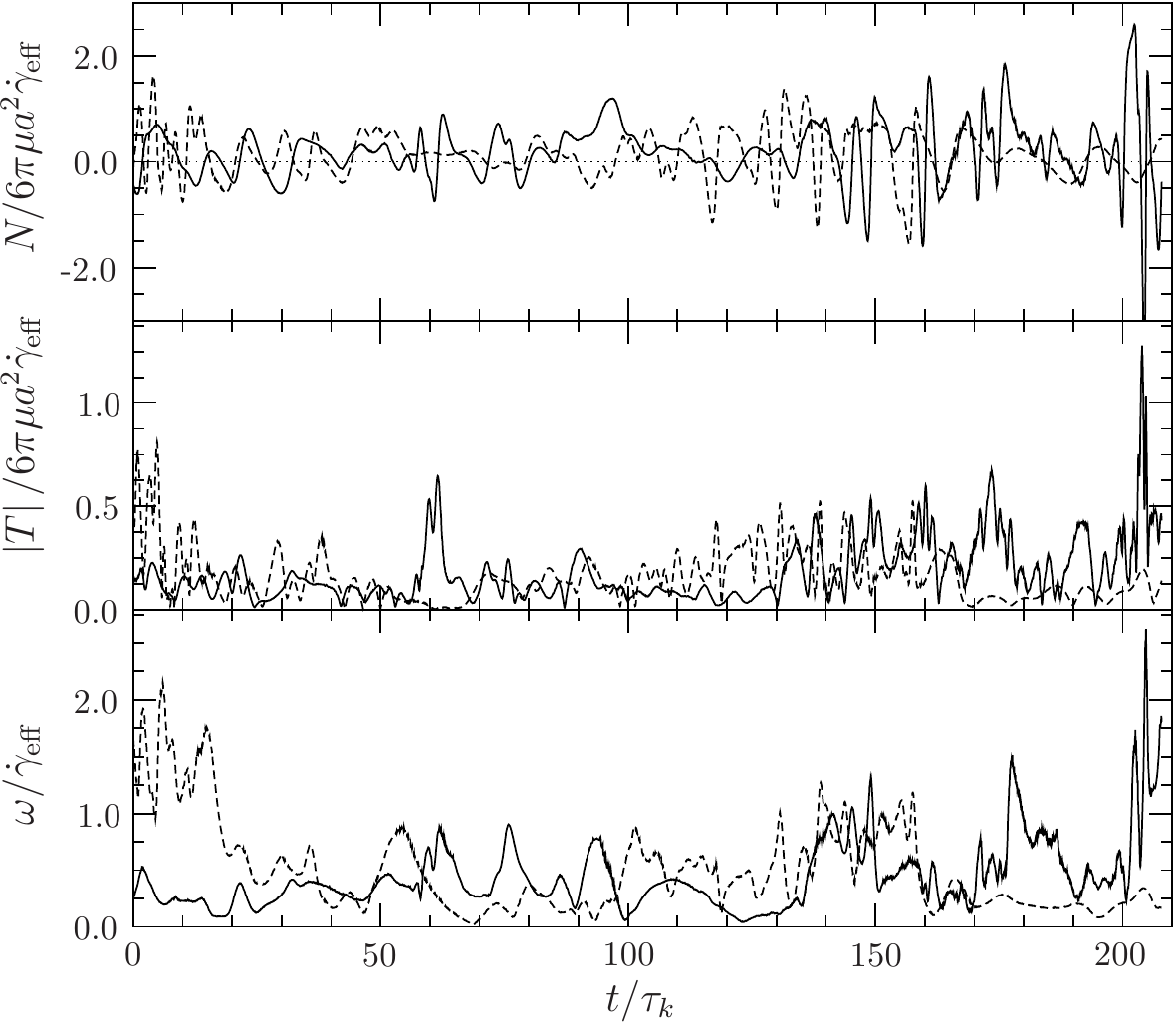}}
  \caption{Time series of normal force $N$, transverse force $T$ and angular velocity $\omega$ for a doublet along two different paths (continuous and dashed line: same paths as in figure~\ref{fig:trajectories1}).}
\label{fig:trajectories2_doubl}
\end{figure}

As shown by the plot of angular velocity in
figure~\ref{fig:trajectories2_doubl}, a doublet rotates irregularly in
the turbulent field and hence changes continuously its orientation
with respect to the velocity gradient. As a consequence, the normal
stress $N$ alternates conditions of traction and compression, changing
in sign at intervals that are on the average around 5$\tau_k$. The
transverse force $T$ also oscillates, and changes alternatively
direction, although this last feature cannot be noticed by the figure,
where only the modulus of $T$ is plotted. The
peaks of $T$ normally occur when the angular velocity is large and
have smaller intensity in comparison to the peaks of $N$.

\begin{figure}%[!h]
  \centerline{\includegraphics[width=\textwidth]{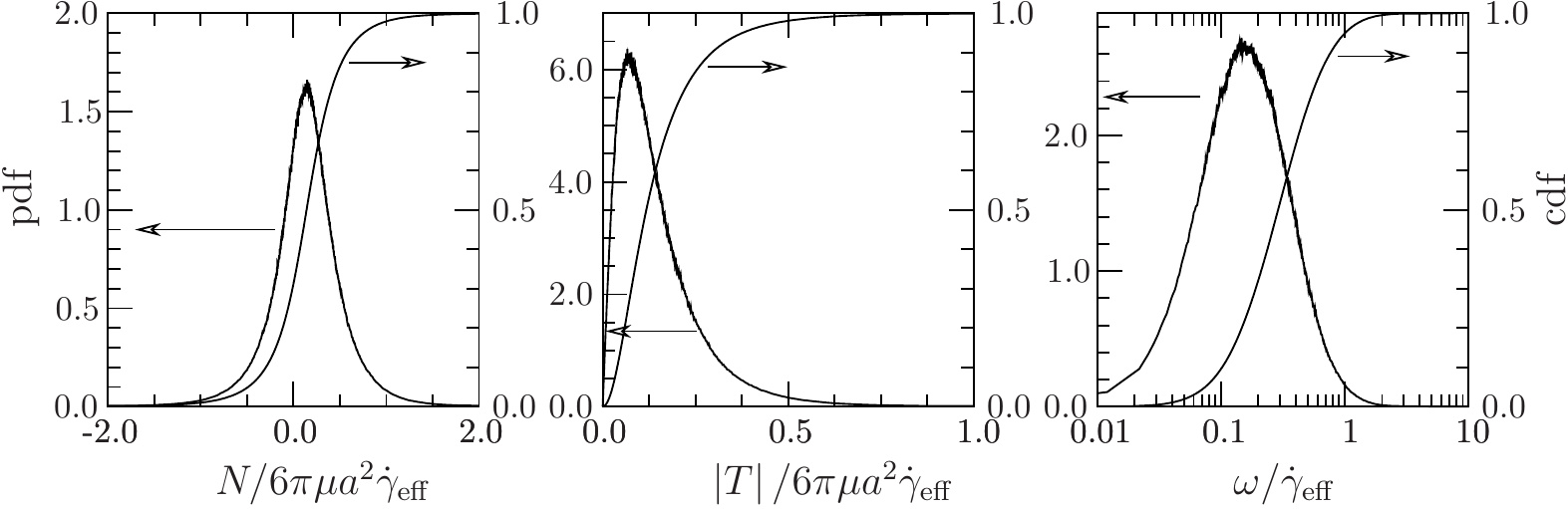}}
  \caption{Probability density functions (pdf, values on the left
    axes) and cumulative distribution functions (cdf, values on the
    right axes) of the intermonomer normal force $N$ and transverse
    force $T$, and of the angular velocity $\omega$ of doublets in the
    turbulent flow field.}
\label{fig:dbl_properties}
\end{figure}
 
\begin{figure}%[!ht]
\begin{center}
\includegraphics[width=0.7\textwidth]{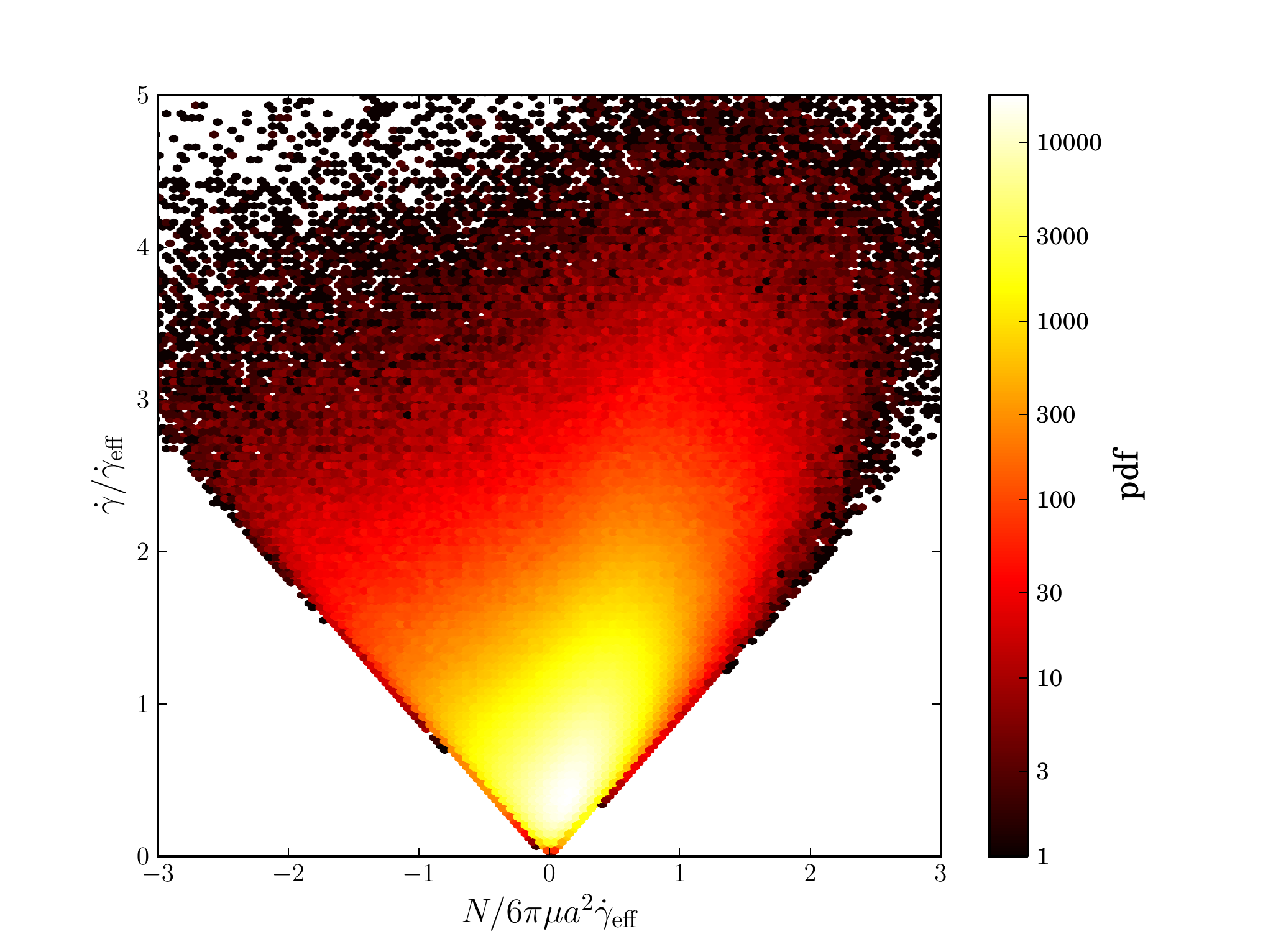}
\caption{Joint shear rate - normal stress distribution for doublets in
  the turbulent flow. Note that the scale of the density function is
  logarithmic in order to better capture the region where the pdf is
  very small.}
\label{fig:jointNG_doublet}
\end{center}
\end{figure}

The angular velocity $\omega$ of a doublet is determined above all by
the shear rate $\dot\gamma$ of the flow field and indeed the
probability distribution for $\omega$ is lognormal as for the shear
rate (figure~\ref{fig:dbl_properties}). The pdf of the normal stress
includes both positive and negative components and is described quite
well by an exponential law decaying as $e^{-\alpha
  \left|x-\bar{x}\right|}$ { or, for large values of $x$, as
  $e^{-\alpha x}$}.  { As also observed by \citet{Derksen_2008},}
tensile stresses prevail over compressive ones, and the probability
that normal stresses are compressive, i.e. negative, is only 31\%, as
shown by the cumulative density function. This fact is not an
inhomogenous sampling of the flow by the aggregate, since in our work
particle inertia is neglected. Rather it may be a consequence of the
preferential alignment with vorticity shown by elongated particles in
turbulent flows \citep{Pumir_2011,Parsa_2012}.  The skewness of the
distribution toward positive normal stresses is also apparent in the
joint shear rate-normal stress distribution of
figure~\ref{fig:jointNG_doublet}. The density function is enclosed by
the two theoretical limits of largest tensile and compressive
stresses, which are $\left| N \right| / (6 \pi \mu a^2 \dot\gamma_{\rm
  eff}) \leq (h_1+h_2)/12 (\dot\gamma / \dot\gamma_{\rm eff})$,
according to the solution by Nir and Acrivos (1973). { It is worth
  observing that, since $(h_1+h_2)/12 = 1.019 \approx 1$, the term $(6
  \pi \mu a^2 \dot\gamma)$ is a good approximation of the largest
  tensile force that should be expected on a doublet immersed in a
  flow field of strength $\dot\gamma$.}

\begin{figure}
\begin{center}
\includegraphics[width=0.6\textwidth]{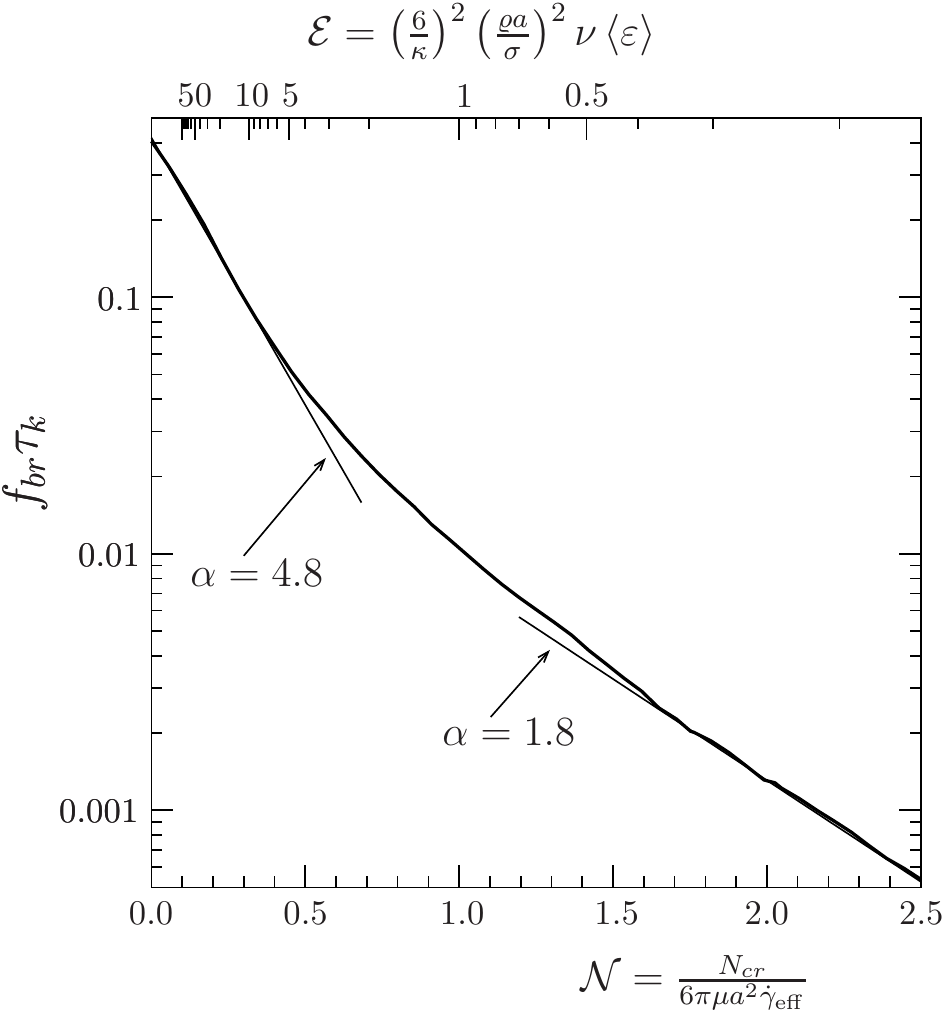} 
\caption{Breakup frequency of a doublet as a function of critical
  stress and average turbulence dissipation rate. The two
  asymptotic regimes of fast and slow breakup are shown by the
  straight lines.}
\label{fig:brfreqdoublet}
\end{center}
\end{figure}

The breakup frequency was estimated by { applying the first-passage
  time analysis described in section~\ref{sec:break_frequency}} on the
full set of the time series of the normal stress and is plotted in
figure~\ref{fig:brfreqdoublet}. It was evaluated for different values
of the variable ${\cal N} = N_{cr}/(6 \pi \mu a^2 \dot\gamma_{\rm
  eff})$ or ${\cal E} = 1/{\cal N}^2 = (6/\kappa)^2 (\varrho a /
\sigma)^2 \nu \left<\varepsilon\right>$, where $N_{cr}$ is related to
surface energy by eq.~(\ref{eq:pulloff}) and $\dot\gamma_{\rm eff}$ to
$\left<\varepsilon\right>$ by eq.~(\ref{eq:eff_shear_rate}). { The
  numerator of ${\cal N}$, $N_{cr}$, is the cohesive strength of the
  bond, whereas the denominator, as observed before, estimates the
  largest tensile stress acting on the bond at the effective shear
  rate $\dot\gamma_{\rm eff}$. As it is the ratio of these two
  variables that determines $\cal N$ and governs breakup, the same
  effect can be achieved by acting on either of these
  variables. Hence, low values of $\cal N$ are obtained with weak
  bonds or, equivalently, strong turbulence.}  The breakup frequency
decreases monotonically with ${\cal N}$ and, conversely, increases
with ${\cal E}$. Two asymptotic regimes (slow and fast breakup) can be
observed with exponential dependence between frequency and ${\cal
  E}^{-1/2}$:
\begin{equation}
f_{br} \propto \sqrt\frac{\left< \varepsilon \right>}{\nu}
\exp\left(-\frac{\alpha}{\sqrt{\cal E}}\right)\,.
\label{eq:freq_doublet}
\end{equation}

For a given cohesive force the slow breakage regime takes place in the
systems with weaker turbulence (${\cal N} > 1.5$ or ${\cal E} <
0.5$). In this case, by fitting the data, we obtain $\alpha =
1.8$. For ${\cal N} < 0.5$ or ${\cal E} > 5$ we are in the fast
breakage situation, where the fitted constant is now $\alpha=4.8$.

{  For the slow breakup region the exponential dependence 
of eq.~(\ref{eq:freq_doublet}) can be justified considering that the 
breakup rate is equal to the frequency at which the normal force $N$ 
exceeds the pull-off value $N_{cr}$. The normal force is generated by 
the turbulent fluctuations, whose frequency is of order 
$1/\tau_k$. However, only the fluctuations for which $N$ crosses the 
value $N_{cr}$ are effective for breakup. Overall, the probability 
of attaining such a condition is proportional to the pdf of the 
normal stress evaluated in $N_{cr}$, which is a function that decays 
exponentially with $\cal N$. If the statistics of the normal stress 
are preserved within single fluctuations, then the expected breakup 
frequency becomes indeed $f_{br} \propto \frac{1}{\tau_k} e^{-\alpha {\cal N}} = \sqrt{{\left< \varepsilon \right>}/{\nu}} \cdot \exp\left( -\alpha/\sqrt{\cal E}\right)$. }

\section{Stress distribution in cluster-cluster aggregates}
\label{sec:clu-clu}

Aggregates were generated numerically by using a tunable
cluster-cluster (CC) method, capable of producing aggregates made of
spherical primary particles of radius $a$ and prescribed fractal
dimension $D_f$ \citep{Thouy_1994,Filippov_2000b}. According to the
fractal scaling law, the total number of primary particles $p$ in an
element of the population of aggregates is related to its gyration radius $R_g$ by
the following equation:
\begin{equation}
p = k_f \left( \frac{R_g}{a} \right)^{D_f}\,.
\label{eq:fractal_law}
\end{equation}
For our particles the fractal law was satisfied exactly at each step
of the hierarchical generation method (i.e., at the level of trimers,
hexamers, 12-mers, etc\ldots). The presence of a fractal structure in
the generated aggregates was also checked by examination of the slope
of the two-point density-density correlation function, which counts
the average number of primary particles at a distance $r$ from any
primary particle in the cluster \citep{Sorensen_1997}. Most of the
following results concern aggregates made of 384 primary particles
with $D_f=1.9$, $k_f$=1.2 and $R_g/a=20.8$. 

\begin{figure}%[!ht]
\centering\includegraphics[width=0.95\textwidth]{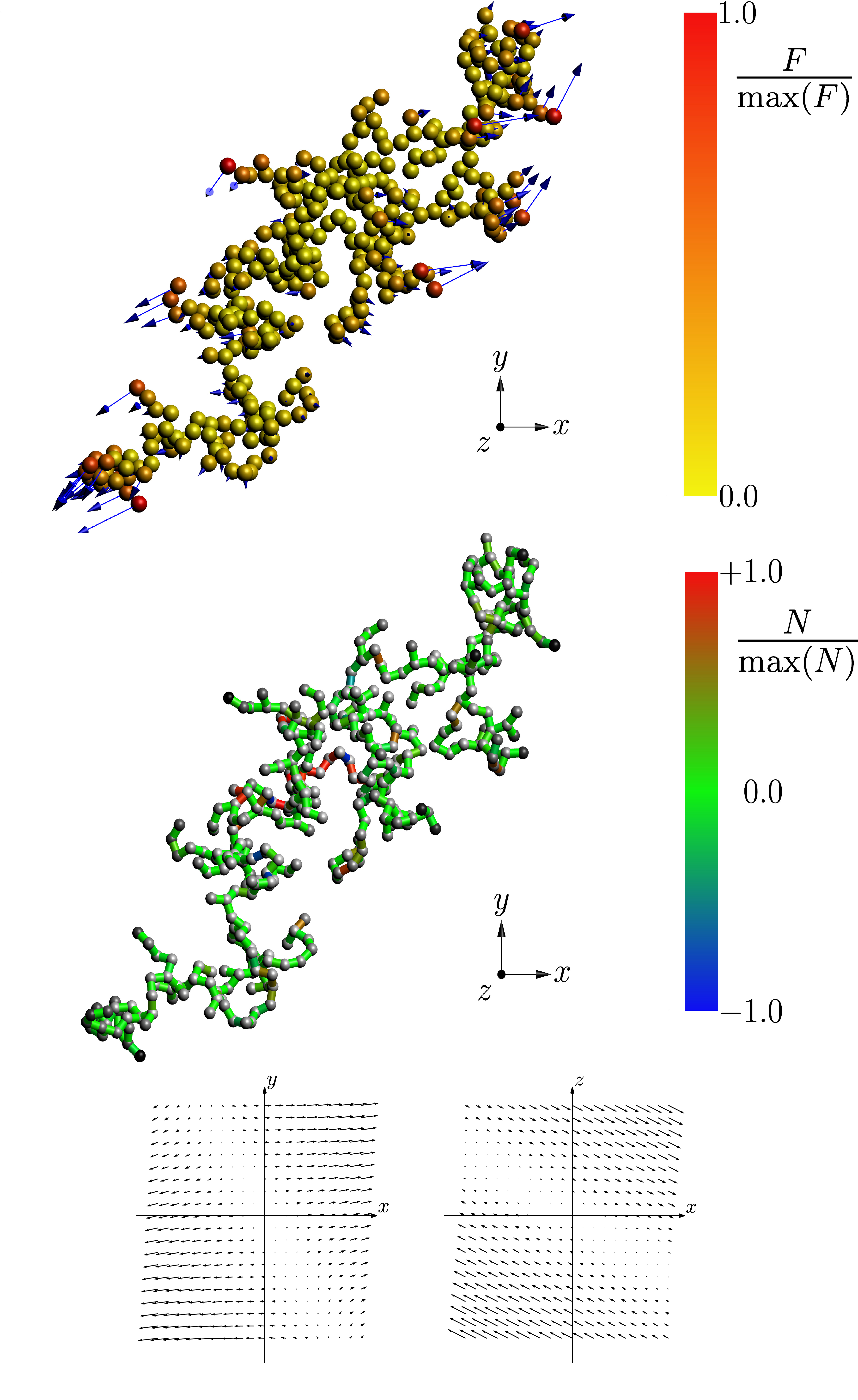}
\caption{Instantaneous hydrodynamic forces (above) and internal normal
  forces (below) for a CC aggregate made of 384 primary particles with
  $D_f=1.9$ and $k_f=1.2$. The instantaneous flow field ${\mathbf u}^\infty$ in the $x-y$ and $x-z$ planes is shown by the vector plots.}
\label{fig:fhydro_cc19}
\end{figure}

The upper part of figure~\ref{fig:fhydro_cc19} shows the forces acting
on the monomers of an aggregate immersed in the turbulent flow field
at a certain time. The forces act in different directions as a
consequence of the three dimensional instantaneous flow field. They
are more intense on the external monomers, { due to the larger slip
  velocity and the smaller hydrodynamic shielding. The centre of mass
  of the aggregate moves with the same velocity of the undisturbed
  flow and hence the slip velocity between primary particles and fluid
  is larger for the outer monomers due to the velocity gradient. In
  addition, when reaching the most exposed external monomers, the
  fluid flow is not hindered or screened by the presence of other
  particles along its path and the drag force is not significantly
  reduced with respect to the case of free draining.}

External forces generate a stress state in the branches of the
aggregate: the lower part of figure~\ref{fig:fhydro_cc19} reports the
normal component of the intermonomer interaction under the same
conditions. Internal forces operate at intermonomer bonds, which have
been evidenced by drawing them as small rods whose colour reflects the
intensity of the normal force. The most stressed bonds, which
correspond to the highest values of tensile normal force can be
identified by the red colour. As already observed in previous papers
\citep{Gastaldi_2011,Vanni_2011}, these critical bonds are located in
the inner region of the cluster and may be quite far from the monomers
with the highest external loads. This is consequence of the structure
of low density aggregates, which are made mostly of chains of primary
particles along which the stress generated by the external forces is
propagated and accumulated.

\begin{figure}%[!ht]
\centering\includegraphics[width=0.8\textwidth]{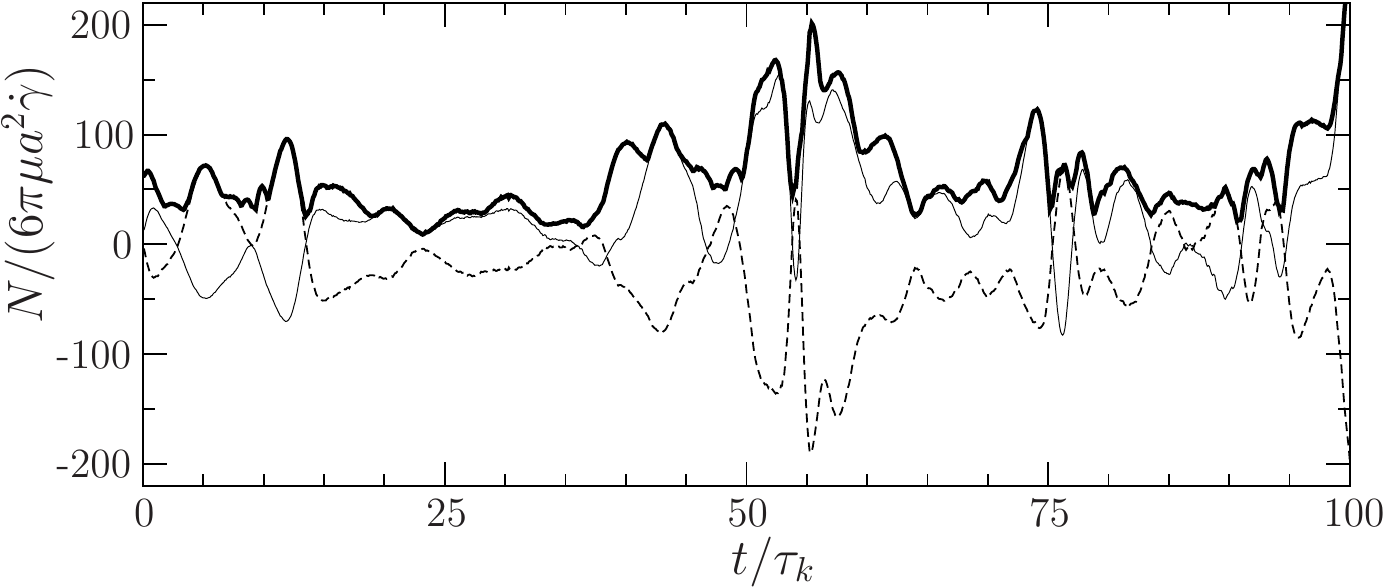}
\caption{Thin lines: fluctuating normal force at two intermonomer
  contacts for the aggregate of figure~\ref{fig:fhydro_cc19}. Thick
  line: maximum value of $N$ over all the bonds of the aggregate,
  $N_{\rm max}$.}
\label{fig:trjCC01}
\end{figure}

Aggregates move and rotate in the turbulent flow field and hence each
intermonomer contact feels a fluctuating force that alternates between
traction and compression. This behaviour is shown in
figure~\ref{fig:trjCC01}, where the normal force component at two
bonds of the aggregate of figure~\ref{fig:fhydro_cc19} is plotted
along time. The figure also reports $N_{\rm max}$, that is, the
largest value of $N$ over all the intermonomer bonds of the cluster at
a given time. Since our aggregates are isostatic, breakup takes place
whenever the tensile stress exceeds the pull-off value at the most
loaded contact of the aggregate. Therefore, it is $N_{\rm max}$ that
should be compared with the threshold value for pull-off in order to determine
the occurrence of breakup.

\begin{figure}%[!ht]
\centering\includegraphics[width=0.6\textwidth]{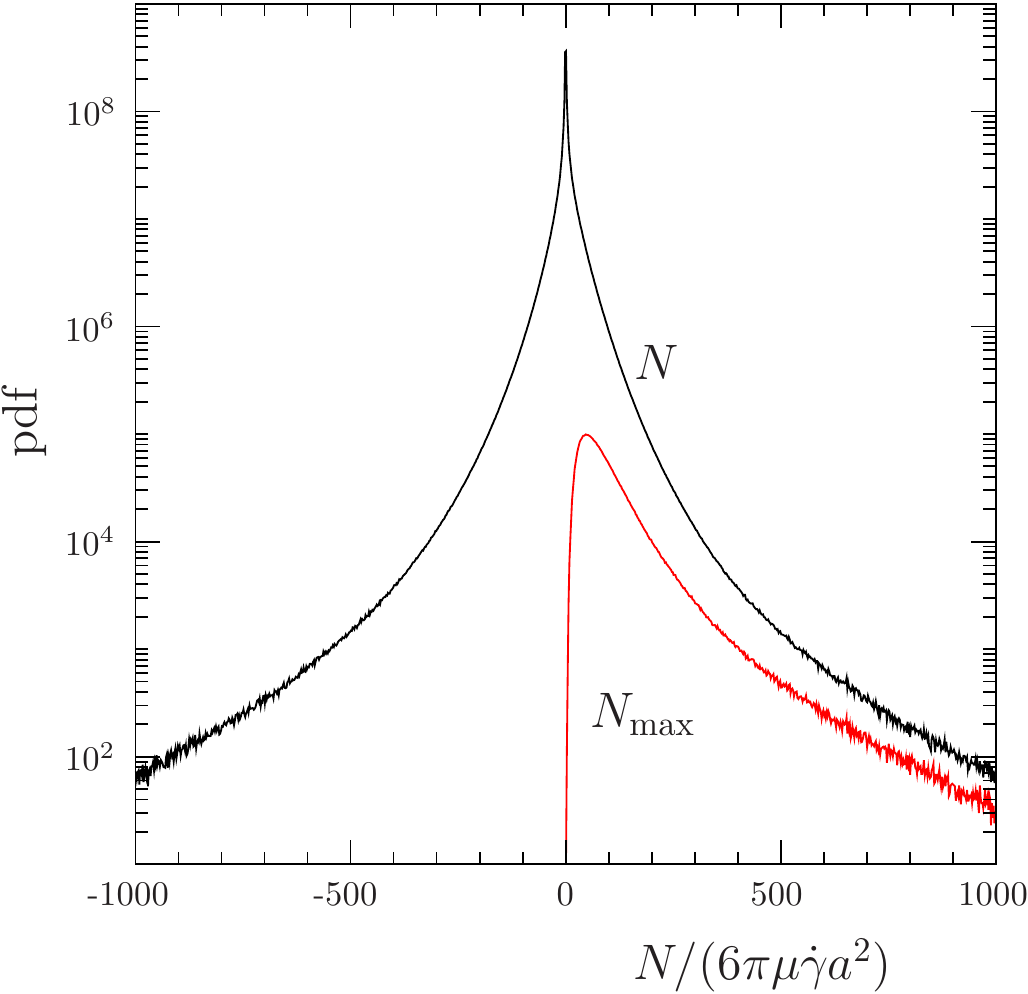}
\caption{Probability density functions for $N$ and $N_{\rm max}$ for
  aggregates with $p=384$, $D_f=1.9$, $k_f=1.2$ in the turbulent
  flow field. Data from a family of 30 different clusters on 3184
  trajectories and sampled at 1393 different times. The area below
  each curve is equal to the number of samples used to build the pdf.}
\label{fig:pdfN_CC19}
\end{figure}

As shown in figure~\ref{fig:pdfN_CC19}, the probability distribution
of the normal intermonomer force follows approximately a stretched
exponential function, $e^{-\left| kx \right|^q}$, with $q\approx
1/3$. The figure includes the data from all the bonds of a family of
30 different aggregates with the same morphology (i.e., same values of
$p$, $D_f$ and $k_f$) on the 3184
turbulent trajectories. As every bond alternates between
traction and compression, the distribution shows positive and negative
values of $N$. Differently from the doublet, where traction prevails,
in this case the distribution is symmetrical and centred at $N=0$. The
reason for the different behaviour lies in the random orientation of
the bonds. While at the level of a single contact the distribution may
be biased toward compression (for example, the thin dashed line of
figure~\ref{fig:trjCC01}) or traction (the thin solid line of the
figure), when all of the bonds of the family of clusters are taken
into account this effect vanishes and the distribution is balanced and
symmetrical. 

\begin{figure}%[!ht]
\centering\includegraphics[width=0.7\textwidth]{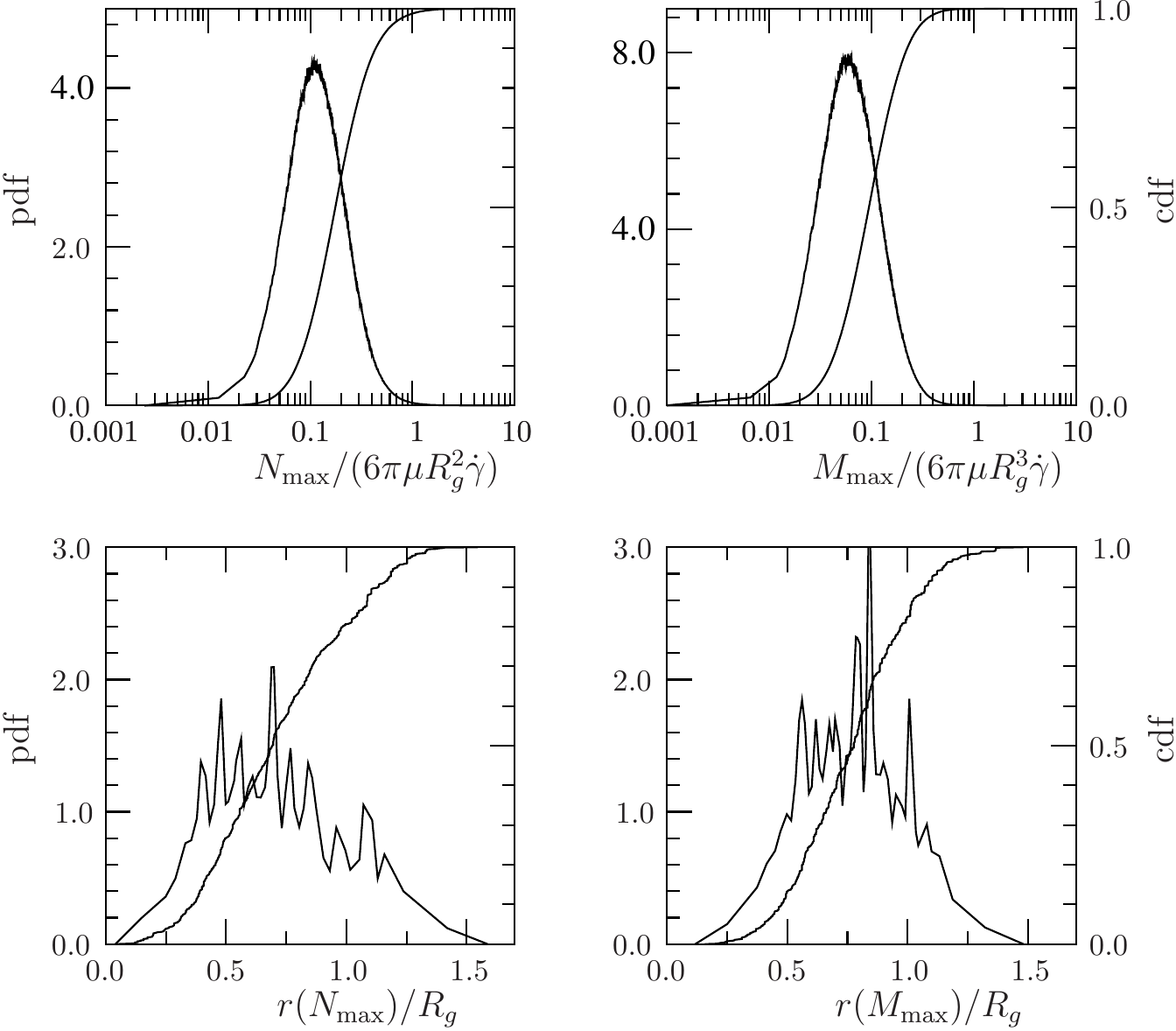}
\caption{Statistical distributions for the population of aggregates
  with $p=384$, $D_f=1.9$, $k_f=1.2$. Above: maximum normal force
  and bending moment; below: radial location of $N_{\rm max}$ and
  $M_{\rm max}$ with respect to the aggregate centre of mass.}
\label{fig:pdf_CC19}
\end{figure}

When only the the maximum normal force $N_{\rm max}$ is considered,
the situation changes substantially. For these aggregates a time
series of $N_{\rm max}$, such as the one shown by the thick line of
figure~\ref{fig:trjCC01}, always mirrors the time series of
$\dot\gamma$. At any time for these large aggregates there exists at
least one bond where the effect of the strain rate is enhanced, due to
favourable location and orientation and, consequently, a peak in
$\dot\gamma$ always implies a simultaneous peak in $N_{\rm max}$. As a
consequence, the pdf of $N_{\rm max}$, too, reflects that of
$\dot\gamma$, and in fact exhibits a similar shape, as apparent by
comparing the distributions shown in figure~\ref{fig:turb_properties}
and figure~\ref{fig:pdf_CC19}.  .  Again, this situation is different
from that of a doublet, where there is smaller correlation between the
instantaneous values of $N_{\rm max}$ (=$N$ in this case) and
$\dot\gamma$, because the orientation of the single bond of the
doublet is not always favourable to the generation of a tensile state
of stress.  { This fact is shown more quantitatively in
  figure~\ref{fig:correl}, where the correlation $\left< \dot\gamma(t)
  N_{\max}(t+\tau) \right>$ is plotted for the doublet and the
  population of cluster-cluster aggregates. While for large aggregates
  the maximum of correlation occurs at the initial time, the doublet
  shows a certain delay in reaching the proper alignment with the flow
  field, which is reflected by a shift in the location of the
  maximum. In addition, the correlation is narrower and decays
  significantly faster than for CC aggregates.

\begin{figure}%[!ht]
\centering\includegraphics[width=0.65\textwidth]{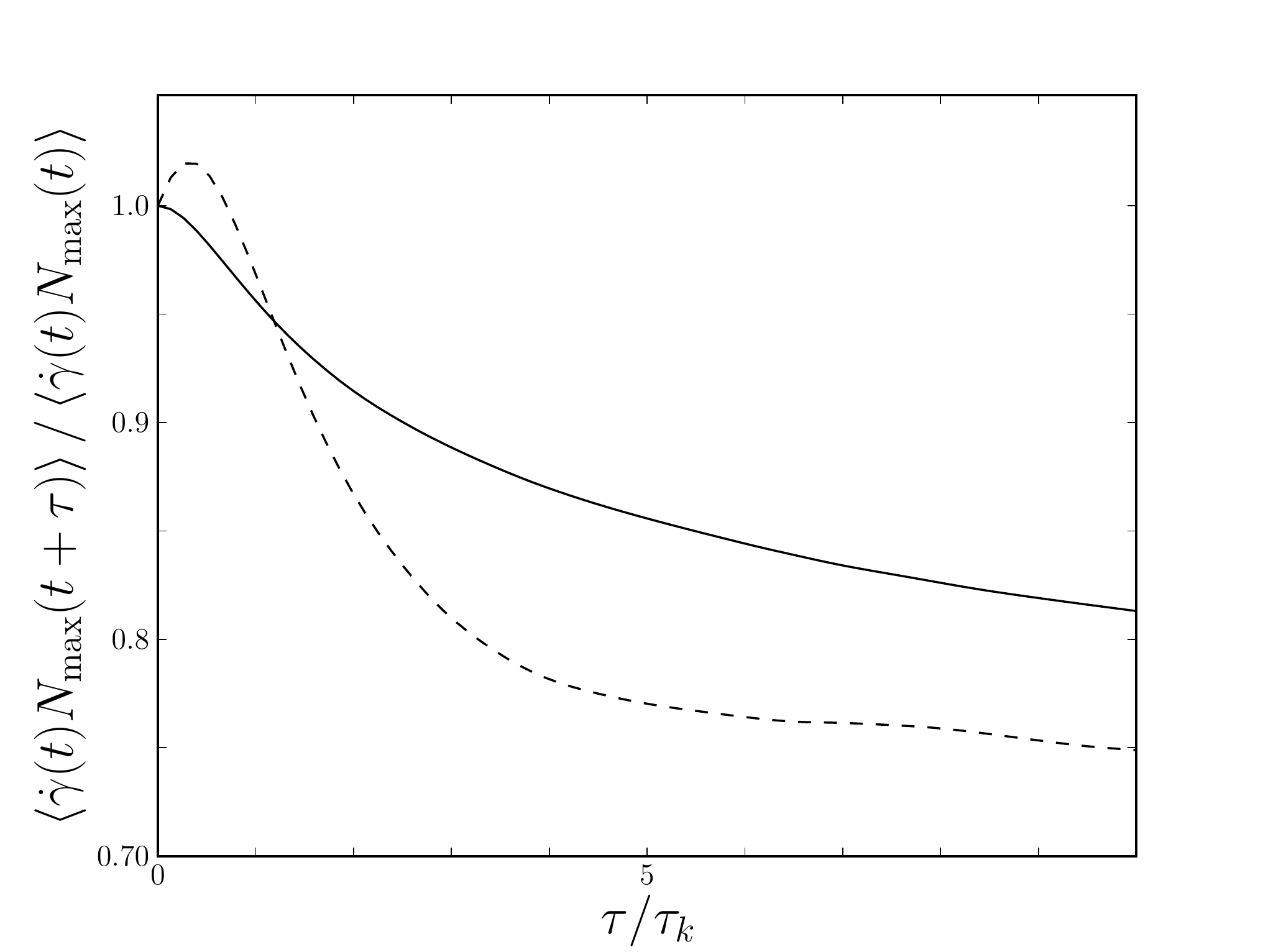}
\caption{Correlation between instantaneous shear rate and maximum
  stress for doublets (dashed line) and the population of
  cluster-cluster aggregates (continuous line).}
\label{fig:correl}
\end{figure}
}

The restructuring of aggregates made of spherical monomers is normally
determined by the bending moment { $M$ and, in particular, by the
  distribution of $M_{\rm max}$, the largest value of $M$ over all the
  bonds of the aggregate}. The behaviour of $M_{\rm max}$ and $N_{\rm
  max}$ is very similar, as visible in figure~\ref{fig:pdf_CC19}, and
both reproduce that of the shear rate.

Figure~\ref{fig:pdf_CC19} also plots the statistical distribution of
the radial location at which the maximum stress (either $N_{\rm max}$
or $M_{\rm max}$) occurs, i.e., the region where rupture or
restructuring are more likely to initiate. Due to the complex shape of
the clusters, the spatial region around the centre of mass is often
empty and the most stressed contacts, generated by the mechanism of
accumulation of internal forces along filaments, are normally located in
a region which is approximately midway between the centre of mass and
the periphery of the aggregate. For the normal force the highest
stress is usually in the interval 0.4 $< r/R_g <$ 0.9; the most
intense bending moments, which have a narrower distribution, are
located in 0.5 $< r/R_g <$ 0.8. Table~\ref{tab:pdf_CC19}
reports average values and standard deviations for all these
distributions, based on the population of clusters made by 384 primary
particles. Data are scaled by the radius of gyration because, as shown
below, this is the relevant length scale for aggregate stress statistics.

\begin{table}
\caption{Average and standard deviation of the stress distributions for 
the population of aggregates with $p=384$, $D_f=1.9$, $k_f=1.2$.}
\label{tab:pdf_CC19}
\begin{center}
\begin{tabular}{ccc}
                                            & average &  std dev \\
  $N_{\rm max}/6\pi\mu R_g^2 \dot{\gamma}_{\rm eff}$  &  0.231  &  0.217   \\
  $M_{\rm max}/6\pi\mu R_g^3 \dot{\gamma}_{\rm eff}$  &  0.117  &  0.088   \\
  $r(N_{\rm max})/ R_g$                     &  0.712  &  0.288   \\
  $r(M_{\rm max})/ R_g$                     &  0.769  &  0.230  
\end{tabular}
\end{center}
\end{table}

\begin{figure}%[!ht]
\centering\includegraphics[width=\textwidth]{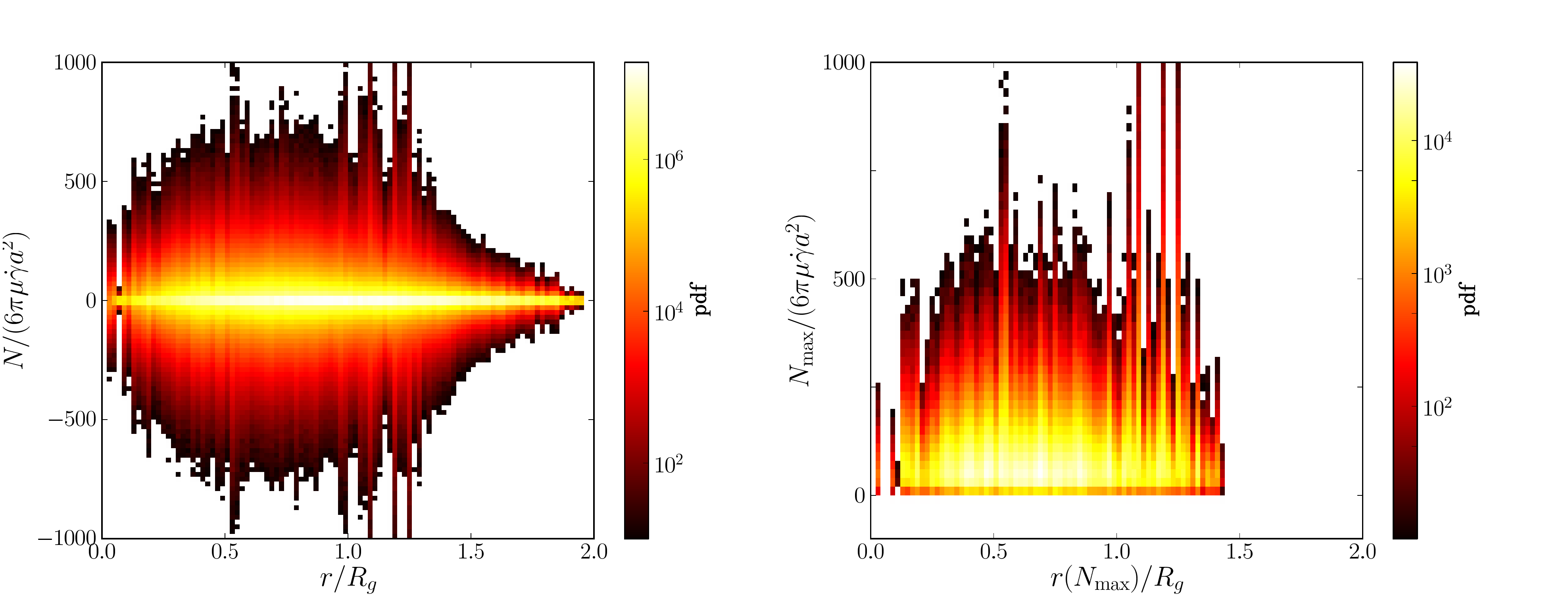}
\caption{Left: joint pdf between normal force and location. Right:
  joint pdf between maximum instantaneous force and
  location. Aggregates with $p=384$, $D_f=1.9$, $k_f=1.2$; data from
  the simulation of 3184 trajectories with 30 different aggregate
  realisations. The scale of the density function is logarithmnic 
  to better describe the region of low pdf.}
\label{fig:stressmax_r02}
\end{figure}

The joint distribution of contact force $N$ and contact location $r$
is shown in the left graph of figure~\ref{fig:stressmax_r02}. It
further confirms that traction and compression are equally probable
and the distribution is symmetric with respect to $N=0$. The strength
of $N$ on most bonds is very small, while the occurrence of large
values of normal force, either in compression or traction, is quite
uncommon. Surprisingly, even when the distribution of 
$N_{\rm max}$ is considered, most conditions still correspond to very
small values of the normal force, while high values of $N_{\rm max}$
occur only at few favourable locations and at particular times. By
comparing the distributions of $N$ and $N_{\rm max}$, one can observe
that, within the collected statistics, the instantaneous maximum of
$N$ never occurs in the outer shell of the aggregate ($1.5 < r/R_g <
2.0$). 

\begin{figure}%[!ht]
\centering\includegraphics[width=0.5\textwidth]{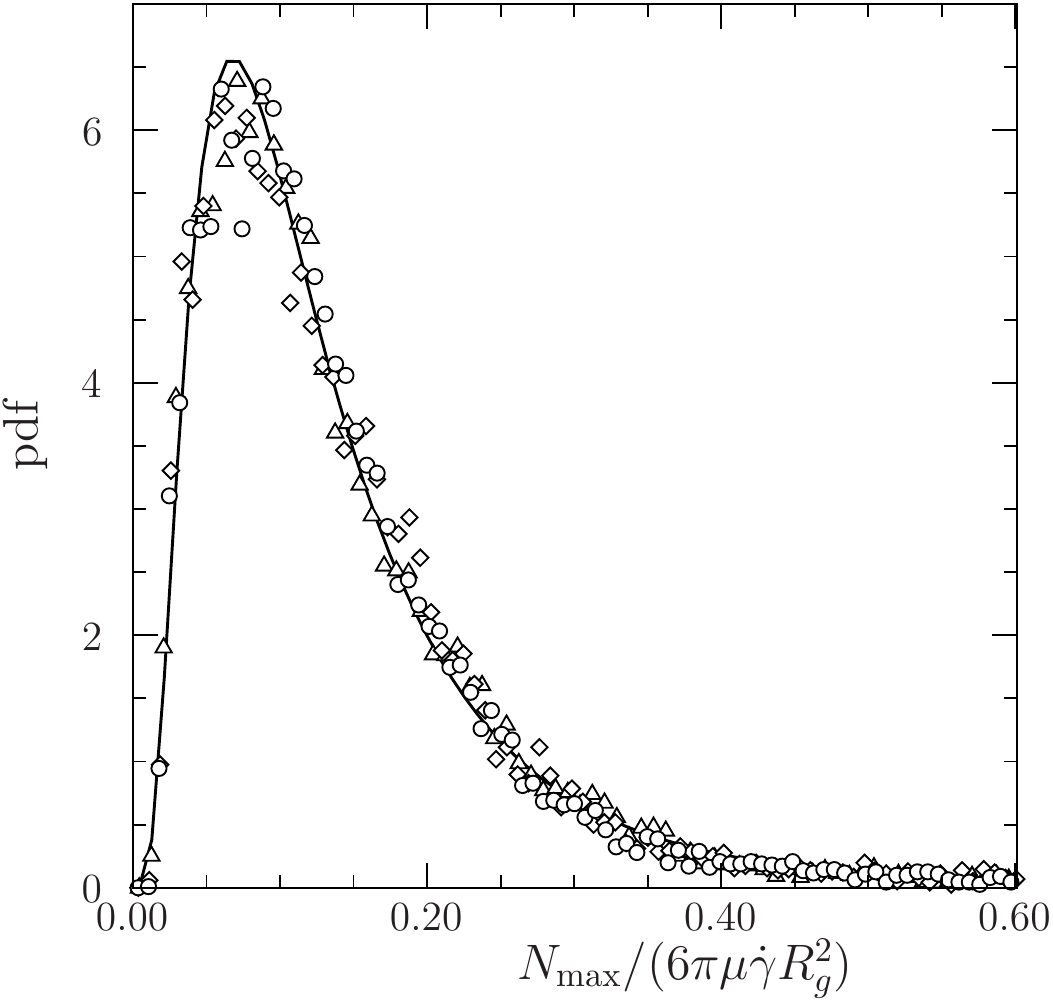}
\caption{Statistical distribution of $N_{\rm max}$ for aggregates with
  $D_f=1.7$. $\circ$: 768 primary particles ($R_g=42.7$); $\diamond$:
  384 primary particles ($R_g=28.4$); $\vartriangle$: 192 primary
  particles ($R_g= 18.9$). For each condition, we examined the response
  of 10 different clusters on 633 trajectories.}
\label{fig:stressmax_r03}
\end{figure}

A previous analysis for pure shear flow suggested that the proper
scaling length for the maximum instantaneous internal forces was the
radius of gyration \citep{Vanni_2011}. In that case, by scaling the 
radial coordinate $r$ by $R_g$, $N_{\rm max}$ by $R_g^2$ and $M_{\rm max}$ by $R_g^3$, the
stress distributions, at least approximately, collapsed on a single
curve, which was independent of particle size, number of monomers and
fractal dimension. This feature is retained by the turbulent flow due to the linearity of the flow
field at the scale of the aggregates. Figure~\ref{fig:stressmax_r03} shows that the statistical
distributions of $N_{\rm max}$ obtained in the studied isotropic
turbulence by aggregates with a number of monomers ranging from 192 to
768 lie on a single curve when scaled by $R_g^2$. The
proportionality by $N_{\rm max}$ and $R_g^2$ can be explained by
considering that the maximum internal stress originates from the
accumulation of drag forces along the chains of monomers that are most
exposed to the flow field. The drag force on the monomers of such
almost unshielded filaments is roughly proportional to $r$, because of
the linear flow surrounding the aggregate, while the length of the
chains is comparable to the radius of gyration.  The moment scales
with $R_g^3$ because of the additional role of the arm of the
force. {  This explanation is confirmed by the fact that in uniform flows, where the drag force on the monomers of the unscreened filaments is approximately constant, $N_{\rm max}$ and $M_{\rm max}$ become proportional to $R_g$ and $R_g^2$, respectively \citep{Gastaldi_2011}. }
It is worth noting that such a scaling is not valid in general
for $N$, $M$ or their average values. The accumulation effect which
determines $N_{\rm max}$ and $M_{\rm max}$ occurs only on the branches
with a favourable (elongated and unshielded) spatial configuration,
whereas for short, wrapped or hydrodynamically screened branches the
effect is negligible.

\begin{figure}%[!ht]
\centering\includegraphics[width=0.6\textwidth]{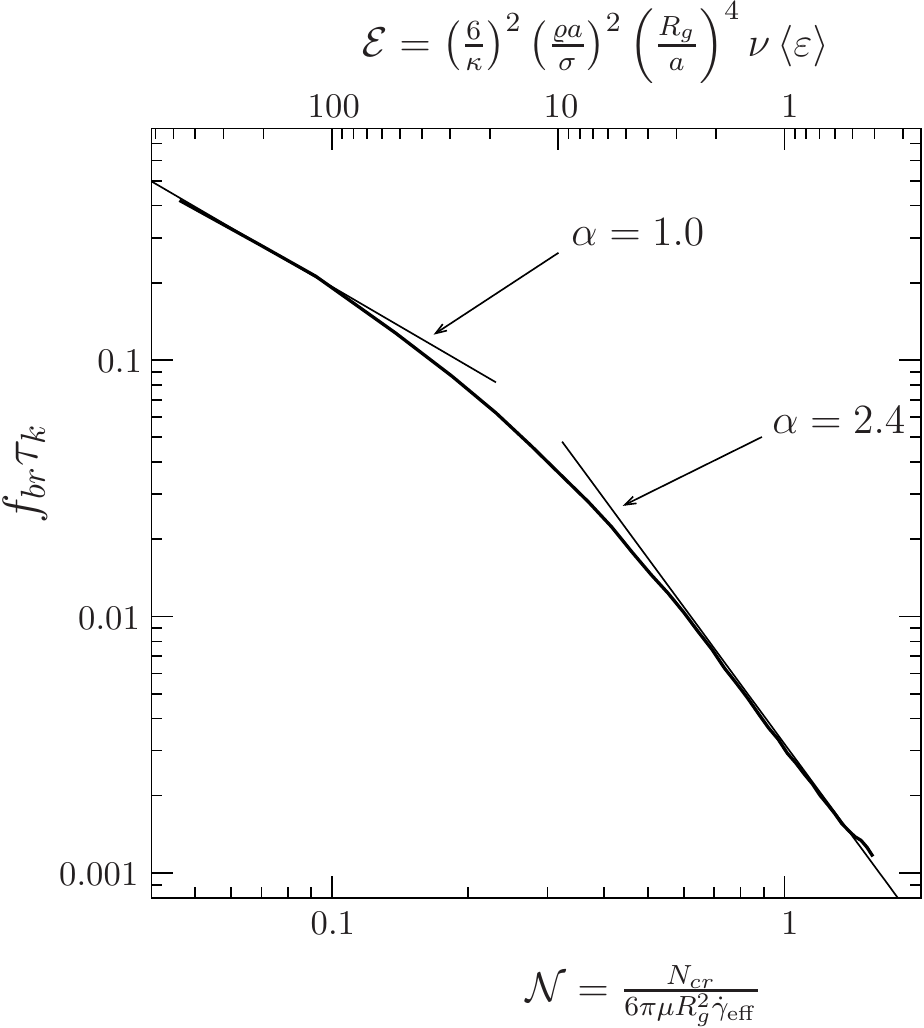}
\caption{Breakage frequency of low-density isostatic cluster-cluster
  aggregates.}
\label{fig:freqN_CC}
\end{figure}

The breakup frequency was estimated by applying first passage time
analysis on the time series of the maximum internal normal force
$N_{\rm max}$. The mean breakup time $\langle \tau \rangle$ was
evaluated by the sequence of diving times obtained by comparing the
profile of $N_{\rm max}$ with the pull-off value $N_{cr}$ for the full
set of 3184 trajectories and 30 realisations of aggregates with
$p=384$, $k_f=1.2$, $D_f=1.9$ and $R_g/a=20.8$. However, the results
can be extended to different situations by taking advantage of the
scaling properties of $N_{\rm max}$ with $R_g^2$, and indeed
figure~\ref{fig:freqN_CC} plots the dimensionless breakup frequency as
a function of ${\cal N} = N_{cr}/(6 \pi \mu \dot\gamma_{\rm eff}
R_g^2)$. Obviously, the scaling is valid only for the conditions in
which it was tested, that is low-density isostatic cluster-cluster
aggregates ($1.7 \leq D_f \leq 2.3$). As turbulence is normally
characterised in terms of the mean dissipation rate $\left<
\varepsilon \right>$ and adhesion in terms of the surface energy
$\sigma$, in figure~\ref{fig:freqN_CC} the breakup frequency is also
plotted as a function of ${\cal E} = 1/{\cal N}^2 = (6/\kappa)^2
(\varrho a / \sigma)^2 (R_g/a)^4 \nu \left<\varepsilon\right>$,
similarly to what was done for the doublet.

Two asymptotic regimes for fast and slow breakup can be identified
that obey power-law relationships:
\begin{equation}
 f_{br} \tau_k \propto \left(1/{\cal N}\right)^{\alpha} = {\cal E}^{\alpha/2}\,,
\label{eq:breakupturb}
\end{equation}
where the power-law exponent $\alpha$ can be fitted by numerical results. For
given physical properties ($\varrho$, $\mu$, $\sigma$) and particle
size ($a$, $R_g$), these regimes also correspond to the limit of
intense and weak turbulence, respectively. In the former one ($\alpha
= 1$, and ${\cal E} \ge 100$, ${\cal N} \le 0.1$), the turbulence is
so intense that most of the fluctuations of the instantaneous strain
rate lead to the rupture of the cluster. The mean breakup time
$\langle \tau \rangle =1/f_{br}$ is comparable to the Kolmogorov time
scale, and hence to the characteristic frequency of the turbulent
fluctuations of the smallest scales. By substituting the expressions
for $\cal E$ and $\tau_k$, the following relationship can be obtained
for this regime:
\begin{equation}
 f_{br}  \propto \frac{\varrho a}{\sigma} 
    \left( \frac{R_g}{a} \right)^2 \left< \varepsilon \right>\,,
\label{eq:breakuptur_fast}    
\end{equation}
showing that the breakup frequency is proportional to the mean
turbulent dissipation rate, to the square of aggregate size and to the
inverse of the cohesive strength.

In the regime of weak turbulence ($\alpha = 2.4$, and ${\cal E} < 4$,
${\cal N} > 0.5$), the breakup frequency is small in comparison to the
frequency of the smallest turbulent fluctuations and only the most
intense peaks of shear rate are capable of breaking an aggregate. The
breakup frequency is:
\begin{equation}
 f_{br} \propto \left(\frac{\varrho a}{\sigma}\right)^{2.4} \left(
 \frac{R_g}{a} \right)^{4.8} \nu^{0.7} \left< \varepsilon
 \right>^{1.7}\,
 \label{eq:slow_breakup_freq}
\end{equation}
and its increase with turbulence strength or particle size is much
steeper than in the condition of fast breakup. Finally a transition
region, where $\alpha$ smoothly changes, separates the two extreme
regimes. All of these regimes occur simultaneously in a
stirred suspension, where particles of different size and regions of
different dissipation rate are present. Consequently, as the variation of the
power-law exponents between the regimes is large, the use of a single
power-law relationship, such as eq.~(\ref{eq:powerlaw_br_frequency}),
to characterise turbulent breakup in dilute suspensions is inadequate.

\begin{figure}%[!ht]
\centering\includegraphics[width=0.6\textwidth]{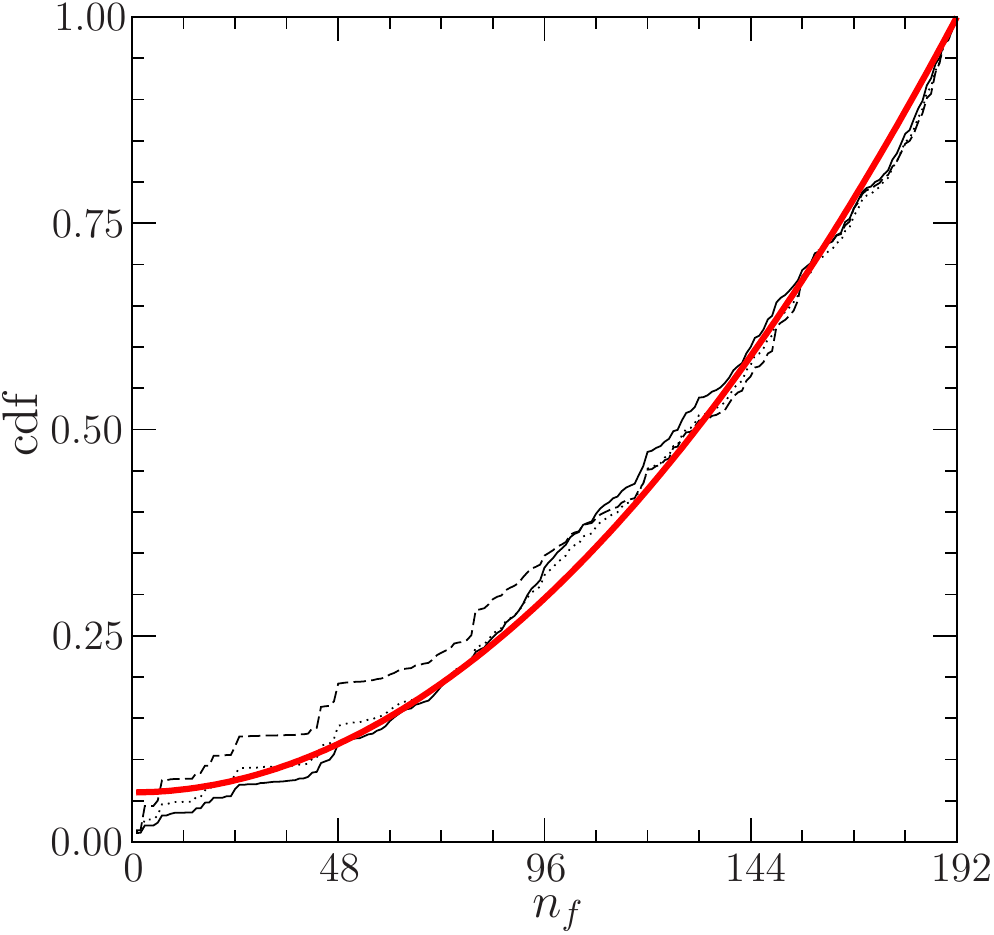}
\caption{Cumulative distribution function for the number of monomers
  in the smaller of the two fragments formed by the breakup of
  384-monomer aggregates ($D_f=1.9$, $k_f=1.2$). Thick red line:
  eq.~(\ref{eq:CDF_fragment}) with $F_0 = 0.06$; thin continuous line:
  ${\cal E}=67$; dotted line: ${\cal E}=33$; dashed line: ${\cal
    E}=10$.}
\label{fig:fragsize}
\end{figure}

The cumulative distribution function for the size of the smaller of
the two fragments formed by the breakup of 384-monomer aggregates is
shown in figure~\ref{fig:fragsize}, where the three curves refer to
values of ${\cal E}$ equal to 10, 33 and 67, respectively. The
cumulative fragment size distribution, $F(n_f)$, can be described
satisfactorily as a function of the number of monomers in the smaller
fragment $n_f$ and the parent aggregate $p$:
\begin{equation}
 \frac{F(n_f)-F_0}{1-F_0} = 4 \left(\frac{n_f}{p} \right)^2 \,.
 \label{eq:CDF_fragment}
\end{equation}
In this relationship $F_0$ characterises the level of erosion, that
is, of fragments formed by 1 or 2 monomers only. The formation of
larger fragments (fragmentation) follows the quadratic function at the
right hand size of this equation. The difference among fragment
distributions in the fast and intermediate regime is quite small and a
single curve with $F_0$ around 5\% is capable of describing
satisfactorily breakup in both regimes. In the slow breakup regime a
larger effect of erosion was observed. In this case, however, the
number of sampled breakup events was much smaller than in the other
two regimes. Although sufficient to evaluate a single parameter such
as the breakup frequency, such a number was too small to give reliable
estimation of the distribution function.

{ The adopted method tracks each cluster up to the start of breakup
  (the failure of the most stressed bond), and gives the breakup
  frequency and the size distribution of the fragments. When the
  formed fragments have moved away from each other, they behave as new
  independent particles, and their breakup frequency can be estimated
  again by first-passage time statistics. The possibility of multiple
  fragmentation events, caused by the failure of additional bonds when
  the fragments are still so close to allow mutual interaction, cannot
  be taken into account by this approach. However, such a situation is
  not common for the considered aggregates. Their breakup normally
  leads to a large and sudden decrease in the maximum stress, because
  it occurs through the rupture of the filament where the effect of
  stress accumulation is largest. An exception could occur if
  different filaments with similar orientation and length, and
  consequently similar levels of stress, are simultaneously present in
  the aggregate, but, due to the low solid density of our clusters,
  this is quite unlikely. }

\begin{figure}%[!h]
  \centerline{\includegraphics[width=0.7\textwidth]{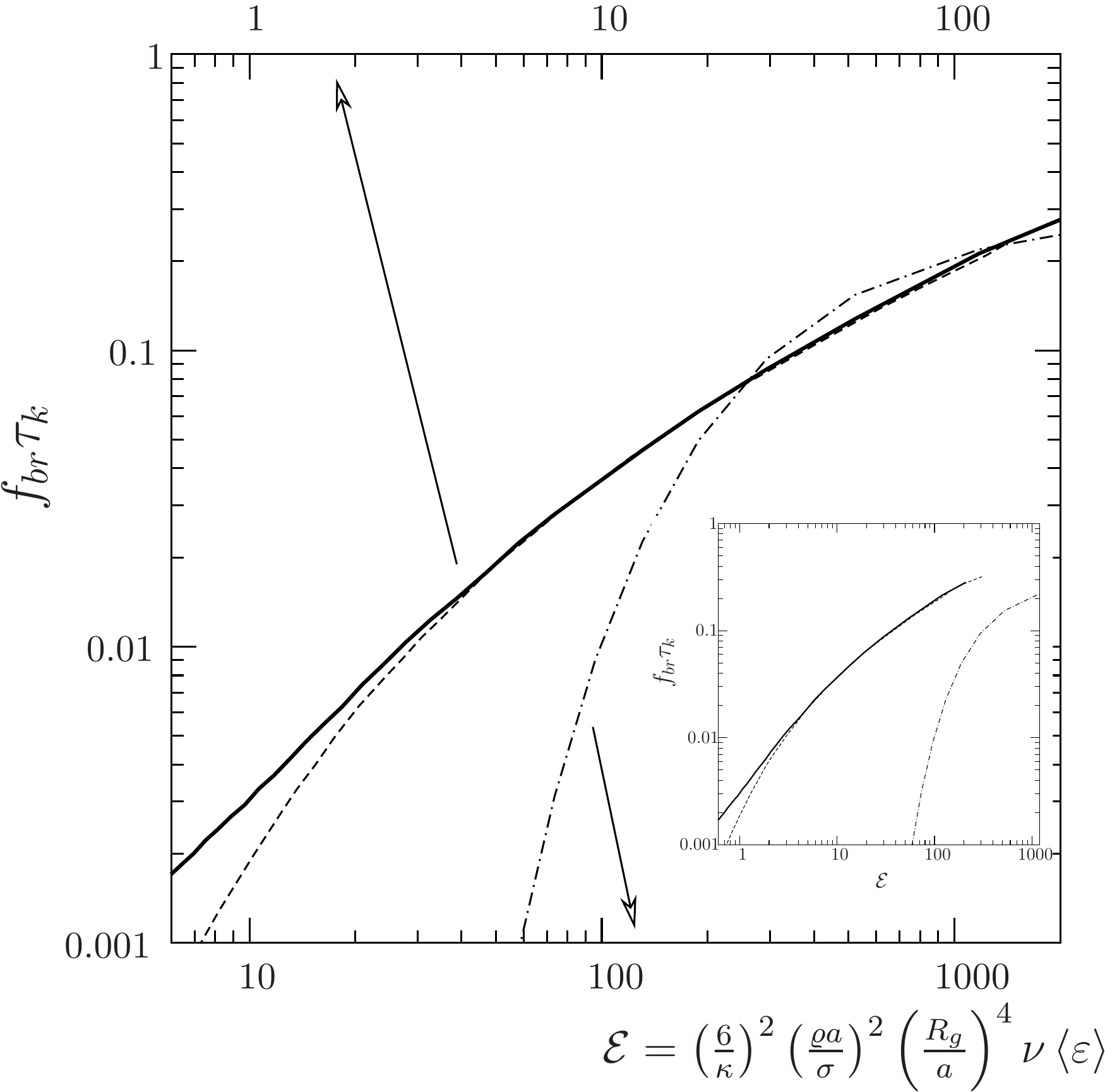}}
  \caption{Comparison of the obtained breakup frequency (continuous
    line) with the results of the methods by \citet{KustersPhD_1991}
    (dot-dashed line) and \citet{Baebler_2012} (dashed line) using
    eq.~(\ref{eq:shear_breakup}) to evaluate the critical shear rate. 
    Two different horizontal scales are used: the lower one for
    the solution by Kusters, and the upper one for our results and the
    method by Ba\"ebler. In the inset, the curves are plotted by one
    horizontal reference scale.}
\label{fig:comp_BaeblerKusters}
\end{figure}

As discussed before, in previous studies the role of internal forces
was neglected and it was assumed that an aggregate breaks up when the
local instantaneous strain rate $\dot\gamma$ exceeds a critical value
$\dot\gamma_{cr}$, which is a function only of the geometry and
cohesive strength of the aggregate. In these studies, rupture does not
depend on the instantaneous orientation of the aggregate with respect
to the local velocity gradient or the distribution of the strain among
the different components of the gradient. Consequently, the critical
shear rate for turbulent breakup can be determined by tests in pure
laminar shear. The predictions of the methods based on this approach
\citep{KustersPhD_1991, Baebler_2012} are compared with our results in
figure~\ref{fig:comp_BaeblerKusters}, using for $\dot\gamma_{cr}$ the
following expression, which is based on a theoretical analysis of the
breakup of isostatic cluster-cluster aggregates in laminar shear flow
\citep[eq. (17)]{Vanni_2011}:
\begin{equation}
 \dot\gamma_{cr} = c \frac{\kappa a \sigma}{\mu R_g^2}\,,
 \label{eq:shear_breakup}
\end{equation}
where $c$ is a coefficient that depends on the morphology of the aggregates.
{  To allow easier comparison with the results of the previous researchers, who related $f_{br}$ to $\left< \varepsilon \right>$, the breakage frequency is here plotted as a function of $\cal E$.}

The expression by Kusters, eq.~(\ref{eq:kusters_br_frequency}), gives
a very steep increase of the breakup frequency with the turbulence
dissipation rate, followed by a sort of saturation of the process at
larger $\left< \varepsilon \right>$. The model assumes a normal
distribution for velocity increments across the aggregate size and
hence for $\dot\gamma$, which is significantly different from the
nearly lognormal one observed in fully developed turbulent flows in
experiments and DNS. { It is probably a consequence of such inadequacy
  the fact that, when applied to the description of processes of
  aggregation-breakup conducted under a wide range of effective shear
  rates, Kusters' equation is often modified by substituting the
  critical shear rate $\dot\gamma_{cr}$, which should depend only on
  particle size, with an empirical function of both size and effective
  shear rate \citep{Fletsch_1999,Marchisio_2006}. }

The first passage time method was proposed by B\"abler and co-authors
(2012), and applied to the same set of turbulent trajectories used for
the present work. In the fast breakup region of $f_{br}\tau_k \approx
1$, where ${\cal N}$ is very small and ${\cal E}$ is large, both
methods predict a linear relationship between $f_{br}$ and $\left<
\varepsilon \right>$. In this regime the turbulent fluctuations are so
much larger than the critical value needed for rupture, that they can
break an aggregate even under unfavourable orientation and thus the
results of the two methods must coincide. Hence, in the regime of fast
breakup, the effect of the orientation is negligible, and the obtained
power-law behaviour, eq.~(\ref{eq:breakuptur_fast}), recovers the
result obtained without accounting for Stokesian dynamics. The
coefficient $c$ of eq.~(\ref{eq:shear_breakup}) has no influence on
the slope of the curve, but only shifts horizontally the solution by
B\"abler and coworkers. In this case its value was set to $c=0.9$ in
order to impose the superposition of the two curves on the basis of
the previous physical reasoning.  The prescribed value is different,
although not largely, from the original one ($c=1.4$) by Vanni and
Gastaldi. Such a difference, however, is probably a consequence of the
use of a different set of clusters and a different method of
extracting statistics between the present work and the paper by Vanni
and Gastaldi.

In the slow breakage region of $f_{br}\tau_k \ll 1$ (characterised by
large values of ${\cal N}$), the difference between the two approaches
becomes relevant and the method by B\"abler and co-authors predicts a
falloff of the breakup rate that is faster than the power-law
prediction of eq.~(\ref{eq:slow_breakup_freq}). In this case the
fluctuations capable of breaking up an aggregate are rare and their
strength does not normally exceed significantly the aggregate
cohesion. In these situations, orientation becomes important for the
outcome of the process in that, by proper alignment of the aggregate
with the flow field, the internal stress at a bond may be enhanced or
dampened.

{ The effect of the Reynolds number was not examined in this work.  It
  is generally believed that the energy dissipation rate goes to a
  finite limit as Reynolds number goes to infinity. This is supported
  by a number of experimental observations, and generally accepted
  from a theoretical point of view
  \citep{Frisch_1995,Sreenivasan_1997}. Moments of the energy
  dissipation depend on the Reynolds number as power laws, with
  scaling exponents that are thought to be universal for isotropic and
  homogeneous turbulence (and independent of Reynolds). However these
  exponents are measured to be extremely small, so a Reynolds
  dependency of the break up rate could well exist, but we expect it
  to be small or vanishing.  }

Real aggregates, with features similar to those of the model
aggregates examined so far (i.e., isostatic clusters of low fractal
dimension) are normally obtained by aggregation of highly destabilised
colloidal suspensions, in which all collisions lead to a permanent
bond and no restructuring of the structures has occurred
\citep{Vanni_2000a}. Since the generation method of such aggregates
proceeds through a sequential addition of smaller clusters to a
growing aggregate without any restructuring, the obtained structure is
isostatic and made mostly of filaments of primary particles, attached
at one end and without internal loops. In these systems the viscous
stresses that cause breakup also cause the restructuring of the
aggregates, by making their structure more compact and
hyperstatic. Hence, the above analysis is only applicable to the
initial steps of a breakup process, when the role of restructuring is
still negligible.

{ The effect of a hyperstatic structure on breakup depends on the
  level of overconstrainment. Highly hyperstatic aggregates show a
  strongly cross-linked structure, in which the hydrodynamic force
  acting on a monomer is distributed on its numerous contacting
  neighbours and discharged locally, instead of being propagated along
  chains of particles. This feature makes the internal stress
  distribution and the mechanism of breakup very different with
  respect to isostatic clusters. Preliminary work based on the
  adoption of a linear stress-strain relationship for contact
  deformations \citep{Vanni_2014} shows indeed that in this case the
  maximum stress is always located at the periphery of the aggregate,
  where the largest hydrodynamic force is applied. The process of
  breakup is more complex in this case, since the failure of the bond
  with the largest stress may give rise to different outcomes: simple
  redistribution of internal stresses because of the redundancy of
  links, direct generation of a small fragment at the periphery of the
  aggregate, start of a sequence of failures of intermonomer links
  that eventually form a crack and split the aggregate in
  fragments. On the contrary, in weakly hyperstatic aggregates, the
  mechanism of transmission and propagation of internal stresses along
  chains of particles is retained. The strength of the internal stress
  for these structures and, consequently, the breakup frequency are
  smaller in comparison to isostatic clusters, since the weakest bonds
  are removed by the restructuring process, but we expect the
  qualitative features concerning spatial distribution of stress, size
  of fragments, breakup regimes to be similar, because of the
  preservation of the filamentous structure. }

\section{Conclusions}
By characterising the stresses generated by homogeneous isotropic
turbulence in rigid aggregates, we estimated theoretically the rate of
turbulent breakup of colloidal aggregates and the size distribution of
the formed fragments. { We examined isostatic aggregates, in which the
  failure of a single contact causes the rupture of the aggregate. It
  is likely that most of the features of such particles are retained
  also by weakly hyperstatic aggregates. On the contrary, highly
  compact and overconstrained clusters have a very different response
  to the hydrodynamic flow field, leading to a more complex mechanism
  of breakup, which may occur by nucleation and propoagation of cracks
  and can not be captured by the present analysis.}

For a given aggregate, the internal stresses depend on the
instantaneous flow field around the aggregate and its orientation with
respect to the flow, which, in turn, depends on the Lagrangian history
of the particle. As a consequence, the translational and rotational
motion of the aggregate needs to be tracked. We considered two classes
of particles, namely doublets and low-density large isostatic
aggregates. Doublets show the highest sensitivity to the effect of
orientation on breakup, as it is the orientation that determines if
their intermonomer contact is under traction or compression. Hence,
since the failure of a contact is caused by a tensile stress, breakup
can only occur if the doublet shows a favourable alignment with
respect to the flow field. On the contrary, in large and highly
disordered aggregates there are contacts under traction for any
spatial configuration. Therefore, these aggregates can be broken under
most orientations, provided that the strength of the strain rate is
large enough. Clearly in this case the effect of orientation is much
smaller, but nevertheless still significant, especially for conditions
of slow breakup.

Asymptotic regimes with limiting laws have been found for the two
cases in which the time-averaged stresses induced in the solid by
turbulence are much larger or much smaller than the cohesive strength
of the aggregate, corresponding to the conditions of fast and slow
breakup, respectively. For large aggregates these limiting
relationships have a power law form and become particularly simple for
the fast breakup region, where the breakup frequency is proportional
to the mean turbulent dissipation rate, the square of aggregate size
and the inverse of the cohesive strength. For such aggregates it has
been confirmed that the proper scaling length for maximum stress and
breakup is the radius of gyration. The cumulative fragment
distribution function is nearly independent of the mean turbulent
dissipation and { can be approximated by the sum of a small erosive
  component and a term that is quadratic with respect to fragment
  size}.

\begin{acknowledgments}
Support from the EU COST Action MP0806 ''Particles in
Turbulence'' for A.L. and M.V. is acknowledged. We thank CINECA
Super Computing Center (Italy), for hosting iCFD database and for
technical support.
\end{acknowledgments}

\bibliographystyle{jfm}

\bibliography{biblio_agg}

\begin{thebibliography}{63}
\expandafter\ifx\csname natexlab\endcsname\relax\def\natexlab#1{#1}\fi

\bibitem[Adler \& Mills(1979)]{Adler_1979a}
{\sc Adler, P.M. \& Mills, P.M.} 1979 Motion and rupture of a porous sphere in
  a linear flow field. {\em Journal of Rheology\/} {\bf 23}, 25--38.

\bibitem[Anderson {\em et~al.\/}(1999)Anderson, Bai, Bischof, Blackford,
  Demmel, Dongarra, Du~Croz, Greenbaum, Hammarling, McKenney \&
  Sorensen]{Anderson_1999}
{\sc Anderson, E., Bai, Z., Bischof, C., Blackford, S., Demmel, J., Dongarra,
  J., Du~Croz, J., Greenbaum, A., Hammarling, S., McKenney, A. \& Sorensen, D.}
  1999 {\em LAPACK Users' Guide\/}, 3rd edn. Philadelphia: SIAM.

\bibitem[B\"abler {\em et~al.\/}(2012)B\"abler, Biferale \&
  Lanotte]{Baebler_2012}
{\sc B\"abler, M.U., Biferale, L. \& Lanotte, A.S.} 2012 Breakup of small
  aggregates driven by turbulent hydrodynamical stress. {\em Physical Review
  E\/} {\bf 85}, 025301.

\bibitem[B\"abler {\em et~al.\/}(2008)B\"abler, Morbidelli \&
  Baldyga]{Baebler_2008a}
{\sc B\"abler, M.U., Morbidelli, M. \& Baldyga, J.} 2008 Modelling the breakup
  of solid aggregates in turbulent flows. {\em Journal of Fluid Mechanics\/}
  {\bf 612}, 261--289.

\bibitem[Bache(2004)]{Bache_2004}
{\sc Bache, D.H.} 2004 Floc rupture and turbulence: a framework for analysis.
  {\em Chemical Engineering Science\/} {\bf 59}, 2521--2534.

\bibitem[Bec {\em et~al.\/}(2010{\natexlab{{\em a\/}}})Bec, Biferale, Cencini,
  Lanotte \& Toschi]{Bec_2010a}
{\sc Bec, J., Biferale, L., Cencini, M., Lanotte, A.S. \& Toschi, F.}
  2010{\natexlab{{\em a\/}}} Intermitency in the velocity distribution of heavy
  particles in turbulence. {\em Journal of Fluid Mechanics\/} {\bf 646},
  527--536.

\bibitem[Bec {\em et~al.\/}(2010{\natexlab{{\em b\/}}})Bec, Biferale, Lanotte,
  Scagliarini \& Toschi]{Bec_2010b}
{\sc Bec, J., Biferale, L., Lanotte, A.S., Scagliarini, A. \& Toschi, F.}
  2010{\natexlab{{\em b\/}}} Turbulent pair dispersion of inertial particles.
  {\em Journal of Fluid Mechanics\/} {\bf 645}, 497--528.

\bibitem[Becker {\em et~al.\/}(2009)Becker, Schlauch, Behr \&
  Briesen]{Becker_2009}
{\sc Becker, V., Schlauch, E., Behr, M. \& Briesen, H.} 2009 Restructuring of
  colloidal aggregates in shear flows and limitations of the free-draining
  approximation. {\em Journal of Colloid and Interface Science\/} {\bf 339},
  362--372.

\bibitem[Bossis {\em et~al.\/}(1991)Bossis, Meunier \& Brady]{Bossis_1991}
{\sc Bossis, G., Meunier, A. \& Brady, J.F.} 1991 Hydrodynamic stress on
  fractal aggregates of spheres. {\em The Journal of Chemical Physics\/} {\bf
  94}, 5064--5070.

\bibitem[Brady \& Bossis(1988)]{Brady_1988}
{\sc Brady, J.F. \& Bossis, G.} 1988 Stokesian dynamics. {\em Annual Review of
  Fluid Mechanics\/} {\bf 20}, 111--157.

\bibitem[Bretherton(1962)]{Bretherton_1962}
{\sc Bretherton, F.P.} 1962 The motion of rigid particles in a shear flow at
  low {R}eynolds numbers. {\em Journal of Fluid Mechanics\/} {\bf 14},
  284--304.

\bibitem[Calvert {\em et~al.\/}(2009)Calvert, Ghadiri \& Tweedie]{Calvert_2009}
{\sc Calvert, G., Ghadiri, M. \& Tweedie, R.} 2009 Aerodynamic dispersion of
  cohesive powders: A review of understanding and technology. {\em Advances
  Powder Technology\/} {\bf 20}, 4--16.

\bibitem[Carpick {\em et~al.\/}(1999)Carpick, Ogletree \&
  Salmeron]{Carpick_1999}
{\sc Carpick, R.W., Ogletree, D.F. \& Salmeron, M.} 1999 A general equation for
  fitting contact area and friction vs load measurements. {\em Advances in
  Colloid and Interface Science\/} {\bf 211}, 395--400.

\bibitem[Delichatsios(1975)]{Delichatsios_1975}
{\sc Delichatsios, M.A.} 1975 Model for the breakup rate of spherical drops in
  isotropic turbulent flows. {\em Physics of Fluids\/} {\bf 18}, 622--623.

\bibitem[Delichatsios \& Probstein(1976)]{Delichatsios_1976}
{\sc Delichatsios, M.A. \& Probstein, R.F.} 1976 The effect of coalescence on
  the average drop size in liquid-liquid dispersions. {\em Industrial \&
  Engineering Chemistry Fundamentals\/} {\bf 15}, 134--138.

\bibitem[Derksen(2008)]{Derksen_2008}
{\sc Derksen, J.J.} 2008 Flow-induced forces in sphere doublets. {\em Journal
  of Fluid Mechanics\/} {\bf 608}, 337--356.

\bibitem[Dukhin {\em et~al.\/}(2005)Dukhin, Zhu, Dave \& Yu]{Dukhin_2005}
{\sc Dukhin, S., Zhu, C., Dave, R.N. \& Yu, Q.} 2005 Hydrodynamic fragmentation
  of nanoparticle aggregates at orthokinetic coagulation. {\em Advances in
  Colloid and Interface Science\/} {\bf 114-115}, 119--131.

\bibitem[Durlofsky {\em et~al.\/}(1987)Durlofsky, Brady \&
  Bossis]{Durlofsky_1987}
{\sc Durlofsky, L., Brady, J.F. \& Bossis, G.} 1987 Dynamic simulation of
  hydrodynamically interacting particles. {\em Journal of Fluid Mechanics\/}
  {\bf 180}, 21--49.

\bibitem[Eggersdorfer {\em et~al.\/}(2010)Eggersdorfer, Kadau, Herrmann \&
  Pratsinis]{Eggersdorfer_2010}
{\sc Eggersdorfer, M.L., Kadau, D., Herrmann, H.J. \& Pratsinis, S.E.} 2010
  Fragmentation and restructuring of soft-agglomerates under shear. {\em
  Journal of Colloid and Interface Science\/} {\bf 342}, 261--268.

\bibitem[Fanelli {\em et~al.\/}(2006)Fanelli, Feke \&
  Manas-Zloczower]{Fanelli_2006a}
{\sc Fanelli, M., Feke, D.L. \& Manas-Zloczower, I.} 2006 Prediction of the
  dispersion of particle clusters in the nano-scale - {P}art {I}: steady
  shearing responses. {\em Chemical Engineering Science\/} {\bf 61}, 473--488.

\bibitem[Filippov {\em et~al.\/}(2000)Filippov, Zurita \&
  Rosner]{Filippov_2000b}
{\sc Filippov, A.V., Zurita, M. \& Rosner, D.E.} 2000 Fractal-like aggregates:
  Relation between morphology and physical properties. {\em Journal of Colloid
  and Interface Science\/} {\bf 229}, 261--273.

\bibitem[Flesch {\em et~al.\/}(1999)Flesch, Spicer \& Pratsinis]{Fletsch_1999}
{\sc Flesch, J.C., Spicer, PT. \& Pratsinis, S.E.} 1999 Laminar and turbulent
  shear-induced flocculation of fractal aggregates. {\em A.I.Ch.E. Journal\/}
  {\bf 45}, 1114--1124.

\bibitem[Frisch(1995)]{Frisch_1995}
{\sc Frisch, U.} 1995 {\em Turbulence: the Legacy of A.N. Kolmogorov\/}.
  Cambridge: Cambridge University Press.

\bibitem[Gastaldi \& Vanni(2011)]{Gastaldi_2011}
{\sc Gastaldi, A. \& Vanni, M.} 2011 The distribution of stresses in rigid
  fractal-like aggregates in a uniform flow field. {\em Journal of Colloid and
  Interface Science\/} {\bf 357}, 18--30.

\bibitem[Goldstein {\em et~al.\/}(1983)Goldstein, Poole \&
  Safko]{Goldstein_1983}
{\sc Goldstein, H., Poole, C. \& Safko, J.} 1983 {\em Classical Mechanics\/},
  3rd edn. San Francisco: Addison-Wesley.

\bibitem[Guazzelli \& Morris(2012)]{Guazzelli_2012}
{\sc Guazzelli, E. \& Morris, J.F.} 2012 {\em A Physical Introduction to
  Suspension Dynamics\/}. Cambridge: Cambridge University Press.

\bibitem[Harada {\em et~al.\/}(2006)Harada, Tanaka, Nogami \&
  Sawada]{Harada_2006}
{\sc Harada, S., Tanaka, R., Nogami, H. \& Sawada, M.} 2006 Dependence of
  fragmentation behavior of colloidal aggregates on their fractal structure.
  {\em Journal of Colloid and Interface Science\/} {\bf 301}, 123--129.

\bibitem[Harshe {\em et~al.\/}(2010)Harshe, Ehrl \& Lattuada]{Harsche_2010}
{\sc Harshe, Y.M., Ehrl, L. \& Lattuada, M.} 2010 Hydrodynamic properties of
  rigid fractal aggregates of arbitrary morphology. {\em Journal of Colloid and
  Interface Science\/} {\bf 352}, 87--98.

\bibitem[Harshe \& Lattuada(2012)]{Harsche_2012}
{\sc Harshe, Y.M. \& Lattuada, M.} 2012 Breakage rate of colloidal aggregates
  in shear flow through {S}tokesian dynamics. {\em Langmuir\/} {\bf 28},
  283--292.

\bibitem[Higashitani \& Iimura(1998)]{Higashitani_1998}
{\sc Higashitani, K. \& Iimura, K.} 1998 Two-dimensional simulation of the
  breakup process of aggregates in shear and elongational flows. {\em Journal
  of Colloid and Interface Science\/} {\bf 204}, 320--327.

\bibitem[Higashitani {\em et~al.\/}(2001)Higashitani, Iimura \&
  Sanda]{Higashitani_2001}
{\sc Higashitani, K., Iimura, K. \& Sanda, H.} 2001 Simulation of deformation
  and breakup of large aggregates in flows of viscous fluids. {\em Chemical
  Engineering Science\/} {\bf 56}, 2927--2938.

\bibitem[Horwatt {\em et~al.\/}(1992{\natexlab{{\em a\/}}})Horwatt, Feke \&
  Manas-Zloczower]{Horwatt_1992b}
{\sc Horwatt, S.W., Feke, D.L. \& Manas-Zloczower, I.} 1992{\natexlab{{\em
  a\/}}} The influence of structural heterogeneities on the cohesivity and
  breakup of agglomerates in simple shear flow. {\em Powder Technology\/} {\bf
  72}, 113--119.

\bibitem[Horwatt {\em et~al.\/}(1992{\natexlab{{\em b\/}}})Horwatt,
  Manas-Zloczower \& Feke]{Horwatt_1992a}
{\sc Horwatt, S.W., Manas-Zloczower, I. \& Feke, D.L.} 1992{\natexlab{{\em
  b\/}}} Dispersion behavior of heterogeneous agglomerates at supercritical
  stresses. {\em Chemical Engineering Science\/} {\bf 47}, 1849--1855.

\bibitem[Ichiki {\em et~al.\/}(2008)Ichiki, Kobryn \& Kovalenko]{Ichiki_2008}
{\sc Ichiki, K., Kobryn, A.E. \& Kovalenko, A.} 2008 Targeting transport
  properties in nanofluidics: hydrodynamic interaction among slip surface
  nanoparticles in solution. {\em Journal of Computational and Theoretical
  Nanoscience\/} {\bf 5}, 2004--2021.

\bibitem[Johnson(1985)]{Johnson_1985}
{\sc Johnson, K.L.} 1985 {\em Contact Mechanics\/}. Cambridge: Cambridge
  University Press.

\bibitem[Kobayashi {\em et~al.\/}(1999)Kobayashi, Adachi \&
  Doi]{Kobayashi_1999}
{\sc Kobayashi, M., Adachi, Y. \& Doi, S.} 1999 Breakup of fractal flocs in a
  turbulent flow. {\em Langmuir\/} {\bf 15}, 4351--4356.

\bibitem[Kusters(1991)]{KustersPhD_1991}
{\sc Kusters, K.A.} 1991 On aggregation of small particles in agitated vessels.
  {PhD} in {C}hemical {E}ngineering, Technische Universiteit Eindhoven.

\bibitem[Lu \& Spielman(1985)]{Lu_1985}
{\sc Lu, C.F. \& Spielman, L.A.} 1985 Kinetics of floc breakage and aggregation
  in agitated liquid suspensions. {\em Journal of Colloid and Interface
  Science\/} {\bf 103}, 95--105.

\bibitem[Manas-Zloczower \& Feke(2009)]{Manas-Zloczower_2009}
{\sc Manas-Zloczower, I. \& Feke, D.} 2009 Dispersive mixing of solid
  additives. In {\em Mixing and Compounding of Polymers: Theory and Practice\/}
  (ed. I.~Manas-Zloczower), pp. 183--216. Munich: Hanser.

\bibitem[Marchisio \& Fox(2013)]{Marchisio_2013}
{\sc Marchisio, D.L \& Fox, R.O.} 2013 {\em Computational Models for
  Polydisperse Particulate and Multiphase Systems\/}. Cambridge University
  Press.

\bibitem[Marchisio {\em et~al.\/}(2006)Marchisio, Soos, Sefcik \&
  Morbidelli]{Marchisio_2006}
{\sc Marchisio, D., Soos, M., Sefcik, J. \& Morbidelli, M.} 2006 Role of
  turbulent shear rate distribution in aggregation and breakage processes. {\em
  AIChE Journal\/} {\bf 52}, 158--173.

\bibitem[Maugis(1992)]{Maugis_1992}
{\sc Maugis, D.} 1992 Adhesion of spheres: The {JKR-DMT} transition using a
  {D}ugdale model. {\em Journal of Colloid and Interface Science\/} {\bf 150},
  243--269.

\bibitem[Nir \& Acrivos(1973)]{Nir_1973}
{\sc Nir, A. \& Acrivos, A.} 1973 On the creeping motion of two arbitrary-sized
  touching spheres in a linear shear field. {\em Journal of Fluid Mechanics\/}
  {\bf 59}, 209--223.

\bibitem[Pandya \& Spielman(1983)]{Pandya_1983}
{\sc Pandya, J.D. \& Spielman, L.A.} 1983 Floc breakage in aggregate
  suspensions: effect of agitation rate. {\em Journal of Colloid and Interface
  Science\/} {\bf 38}, 1983--1992.

\bibitem[Parker {\em et~al.\/}(1972)Parker, Kaufmann \& Jenkins]{Parker_1972}
{\sc Parker, D.S., Kaufmann, W.J. \& Jenkins, J.} 1972 Floc breakup in
  turbulent flocculation processes. {\em Proceedings of the American Society of
  Civil Engineers. Journal of the Sanitary Engineering Division.\/} {\bf 98},
  79--99.

\bibitem[Parsa {\em et~al.\/}(2012)Parsa, Calzavarini, Toschi \&
  Voth]{Parsa_2012}
{\sc Parsa, S., Calzavarini, E., Toschi, F. \& Voth, G.A.} 2012 Rotation rate
  of rods in turbulent fluid flow. {\em Physical Review Letters\/} {\bf 109},
  134501.

\bibitem[Potanin(1993)]{Potanin_1993}
{\sc Potanin, A.A.} 1993 On the computer simulation of the deformation and
  breakup of colloidal aggregates in shear flow. {\em Journal of Colloid and
  Interface Science\/} {\bf 157}, 399--410.

\bibitem[Pumir \& Wilkinson(2011)]{Pumir_2011}
{\sc Pumir, A. \& Wilkinson, M.} 2011 Orientation statistics of small particles
  in turbulence. {\em New Journal of Physics\/} {\bf 13}, 093030.

\bibitem[Redner(2001)]{Redner_2001}
{\sc Redner, S.} 2001 {\em A Guide to First Passage Processes\/}. Cambridge:
  Cambridge University Press.

\bibitem[Sanchez~Fellay \& Vanni(2012)]{Sanchez_2012}
{\sc Sanchez~Fellay, L. \& Vanni, M.} 2012 The effect of flow configuration on
  hydrodynamic stresses and dispersion of low density rigid aggregates. {\em
  Journal of Colloid and Interface Science\/} {\bf 388}, 47--55.

\bibitem[Seto {\em et~al.\/}(2011)Seto, Botet \& Briesen]{Seto_2011}
{\sc Seto, R., Botet, R. \& Briesen, H.} 2011 Hydrodynamic stress on small
  colloidal aggregates in shear flow using {S}tokesian dynamics. {\em Physical
  Review E\/} {\bf 84}, 041405.

\bibitem[Sonntag \& Russel(1987)]{Sonntag_1987a}
{\sc Sonntag, R.C. \& Russel, W.B.} 1987 Structure and breakup of flocs
  subjected to fluid stresses. {II}. {T}heory. {\em Journal of Colloid and
  Interface Science\/} {\bf 115}, 378--389.

\bibitem[Sorensen \& Roberts(1997)]{Sorensen_1997}
{\sc Sorensen, C.M. \& Roberts, G.C.} 1997 The prefactor of fractal aggregates.
  {\em Journal of Colloid and Interface Science\/} {\bf 186}, 447--452.

\bibitem[Spicer \& Pratsinis(1996)]{Spicer_1996c}
{\sc Spicer, P.T. \& Pratsinis, S.E.} 1996 Coagulation and fragmentation:
  Universal steady-state particle-size distribution. {\em AIChE Journal\/} {\bf
  42}, 1612--1620.

\bibitem[Sreenivasan \& Antonia(1997)]{Sreenivasan_1997}
{\sc Sreenivasan, K.R. \& Antonia, R.A.} 1997 The phenomenology of small-scale
  turbulence. {\em Annual Review of Fluid Mechanics\/} {\bf 29}, 435--472.

\bibitem[Tadmor(1976)]{Tadmor_1976}
{\sc Tadmor, Z.} 1976 Forces in dispersive mixing. {\em Industrial Engineering
  Chemistry Fundamentals\/} {\bf 15}, 346--348.

\bibitem[Thouy \& Jullien(1994)]{Thouy_1994}
{\sc Thouy, R. \& Jullien, R.} 1994 A cluster-cluster aggregation model with
  tunable fractal dimension. {\em J. Phys. A: Math. Gen.\/} {\bf 27},
  2953--2963.

\bibitem[Vanni(2000{\natexlab{{\em a\/}}})]{Vanni_2000b}
{\sc Vanni, M.} 2000{\natexlab{{\em a\/}}} Approximate population balance
  equations for aggregation-breakage processes. {\em Journal of Colloid and
  Interface Science\/} {\bf 221}, 243--260.

\bibitem[Vanni(2000{\natexlab{{\em b\/}}})]{Vanni_2000a}
{\sc Vanni, M.} 2000{\natexlab{{\em b\/}}} Creeping flow over spherical
  permeable aggregates. {\em Chemical Engineering Science\/} {\bf 55},
  685--698.

\bibitem[Vanni(2014)]{Vanni_2014}
{\sc Vanni, M.} 2014 Internal stresses and breakup of porous aggregates in
  homogeneous isotropic turbulence. {\em ASME 2014 4th Joint US-European Fluids
  Engineering Division Summer Meeting\/} Submitted.

\bibitem[Vanni \& Gastaldi(2011)]{Vanni_2011}
{\sc Vanni, M. \& Gastaldi, A.} 2011 Hydrodynamic forces and critical stresses
  in low-density aggregates under shear flow. {\em Langmuir\/} {\bf 27},
  12822--12833.

\bibitem[Wengeler \& Nirschl(2007)]{Wengeler_2007}
{\sc Wengeler, R. \& Nirschl, H.} 2007 Turbulent hydrodynamic stress induced
  dispersion and fragmentation of nanoscale agglomerates. {\em Journal of
  Colloid and Interface Science\/} {\bf 306}, 262--273.

\bibitem[Zaccone {\em et~al.\/}(2009)Zaccone, Soos, Lattuada, Wu, B\"abler \&
  Morbidelli]{Zaccone_2009}
{\sc Zaccone, A., Soos, M., Lattuada, M., Wu, H., B\"abler, M. \& Morbidelli,
  M.} 2009 Breakup of dense colloidal aggregates under hydrodynamic stresses.
  {\em Physical Review E\/} {\bf 79}, 061401.

\end{thebibliography}

\end{document}